\documentclass{aa}  

\usepackage{xcolor}
\usepackage{graphicx}
\usepackage{txfonts}
\usepackage{multirow}
\usepackage{hyperref}

\definecolor{rusterucci2024_blue_link_color}{HTML}{3B54A3}
\hypersetup{linktoc=page, colorlinks,  
            linkcolor=rusterucci2024_blue_link_color, 
            citecolor=rusterucci2024_blue_link_color, 
            urlcolor=rusterucci2024_blue_link_color,
            bookmarksopen, bookmarksnumbered}

\titlerunning{Kinematics of the distant spur feature of the Sagittarius 
              stream traced by Blue Horizontal Branch stars}

\defcitealias{pristine_dr1}{%
  Martin, Starkenburg \textit{et al.},
  \textcolor{rusterucci2024_blue_link_color}{2024}}

\begin{document}
  \title{A Pristine-UNIONS view on the Galaxy: Kinematics of the 
         distant spur feature of the Sagittarius stream traced by Blue
         Horizontal Branch stars}
  \author{M.~Bayer\inst{1} \and E.~Starkenburg\inst{1}
          \and G.~F.~Thomas\inst{2, 3} \and N.~Martin\inst{4, 5}
          \and A.~Helmi\inst{1} \and A.~Bystr{\"o}m\inst{6}
          \and T.~de~Boer\inst{7} \and E.~Fern{\'a}ndez Alvar\inst{2, 3}
          \and S.~Gwyn\inst{8} \and R.~Ibata\inst{4}
          \and P.~Jablonka\inst{9} \and G.~Kordopatis\inst{10}
          \and T.~Matsuno\inst{11} \and A.~W.~McConnachie\inst{8}
          \and G.~E.~Medina\inst{12, 13}
          \and R.~S{\'a}nchez-Janssen\inst{14} \and F.~Sestito\inst{15}}
  \institute{Kapteyn Astronomical Institute, University of Groningen,
             Landleven 12, NL-9747AD Groningen, the Netherlands\\
             \email{mbayer@astro.rug.nl}
             \and Instituto de Astrof{\'i}sica de Canarias, E-38205 
             La Laguna, Tenerife, Spain
             \and Universidad de La Laguna, Dpto. Astrof{\'i}sica, 
             E-38206 La Laguna, Tenerife, Spain
             \and Universit{\'e} de Strasbourg, Observatoire
             astronomique de Strasbourg, UMR 7550, F-67000 Strasbourg, 
             France
             \and Max-Planck-Institut f{\"u}r Astronomie, 
             K{\"o}nigstuhl 17, D-69117 Heidelberg, Germany
             \and Institute for Astronomy, University of Edinburgh, 
             Royal Observatory, Blackford Hill, Edinburgh EH9 3HJ, UK
             \and Institute for Astronomy, University of Hawaii, 2680
             Woodlawn Drive, Honolulu HI 96822, USA 
             \and NRC Herzberg Astronomy and Astrophysics, 5071 West
             Saanich Road, Victoria, BC, V9E2E7, Canada
             \and Laboratoire d'astrophysique, École Polytechnique 
             Fédérale de Lausanne (EPFL), CH-1290 Sauverny, Switzerland
             \and Universit{\'e} C{\^o}te d'Azur, Observatoire de la
             C{\^o}te d'Azur, CNRS, Laboratoire Lagrange, 06000 Nice,
             France
             \and Astronomisches Rechen-Institut, Zentrum f{\"u}r 
             Astronomie der Universit{\"a}t Heidelberg,
             M{\"o}nchhofstraße 12–14, 69120 Heidelberg, Germany
             \and David A. Dunlap Department of Astronomy \&
             Astrophysics, University of Toronto, 50 St George Street,
             Toronto ON M5S 3H4, Canada
             \and Dunlap Institute for Astronomy \& Astrophysics,
             University of Toronto, 50 St George Street, Toronto, ON M5S
             3H4, Canada
             \and STFC UK Astronomy Technology Centre, Royal 
             Observatory, Blackford Hill, Edinburgh, EH9 3HJ, UK
             \and Centre for Astrophysics Research, Department of 
             Physics, Astronomy and Mathematics, University of 
             Hertfordshire, Hatfield, AL10 9AB, UK}
  \date{Received - XX, XXXX; accepted - XX, XXXX}
  \abstract%  
  {Providing a detailed picture of the Sagittarius (Sgr) stream offers 
   important constraints on the build-up of the Galactic halo as well 
   as its gravitational potential at large radii. While several 
   attempts have been made to model the structure of the Sgr stream, no 
   model has yet been able to match all the features observed for the 
   stream. Moreover, for several of these features, observational 
   characterisation of their properties is rather limited, particularly 
   at large distances.}%
  {The aim of this work is to investigate the kinematics of the Sgr 
   stream outermost spur feature using blue horizontal branch (BHB) 
   stars.}%
  {Candidate BHB stars were selected by combining two approaches; one
   capitalising on Pan-STARRS1 3$\Pi$ \textit{griz} and \textit{u}
   photometry taken as part of UNIONS, the other using Pristine Survey
   \textit{CaHK} and SDSS \textit{ugr} photometry. Follow-up optical
   spectra are obtained using ESO/VLT/FORS2 to confirm their BHB nature
   and obtain line-of-sight (LOS) velocities.}%
  {Of our 25 candidates, 20 stars can be confirmed as bona fide BHB
   stars. Their LOS velocities, together with the 3D positions of these
   stars qualitatively match well with Sgr model predictions and trace 
   the outer apocentre of the trailing arm and its spur feature very 
   nicely. The quantitative offsets that are found between our data and 
   the different models can be used to provide information about the
   Galactic gravitational potential at large distances. We present a
   first, tentative, analysis in this direction, showing that the model
   of \citet{vasiliev_et_al2021} would provide better agreement with our
   observations if the
   enclosed mass of the Milky Way within 100\,kpc 
   were lowered to
   $(5.3\!\pm\!0.4)\!\times\!10^{11}\,\mathrm{M}_\odot%
   \left(\text{versus}\ (5.6\!\pm\!0.4)\!\times\!10^{11}%
   \,\mathrm{M}_\odot\right)$.}%
  {Our selection of BHB stars provides a new view on the outermost 
   structure in 3D positions and LOS velocities of the Sgr debris.}
  \maketitle

  \renewcommand\sectionautorefname{Sect.}
  \renewcommand\subsectionautorefname{Sect.}
  \renewcommand\figureautorefname{Fig.}

  \section{Introduction}

  The Sagittarius dwarf spheroidal galaxy is one of the largest 
  building blocks of the Milky Way (MW) halo
  \citep[e.\,g.,][]{majewski_et_al2003, read_erkal2019, naidu_et_al2020}
  with an estimated, present, total mass inside the remnant of at least 
  $10^9\,\mathrm{M}_\odot$ \citep{ibata_et_al1997}.
  Over the past 30 years since its original discovery 
  \citep{ibata_et_al1994}, it became clear that the stellar stream 
  produced by the infall of this dwarf galaxy onto the MW is wrapping 
  around the Galaxy fully and it is a major component of the 
  Galactic halo
  \citep[e.\,g.,][]{ibata_et_al2001, majewski_et_al2003, helmi2004,
                    johnston_et_al2005, belokurov_et_al2006,
                    belokurov_et_al2014, sesar_et_al2017b,
                    antoja_et_al2020, ibata_et_al2020, ramos_et_al2020,
                    naidu_et_al2020, ramos_et_al2022}. 
  Specifically, the stellar stream is believed to dominate the outer
  Galactic halo at galactocentric radii beyond 25\,kpc 
  \citep{naidu_et_al2020}.
  
  Many intricate features can be seen in the arms of the Sagittarius 
  debris. With the onset of large-area deep photometric surveys as the 
  Sloan Digital Sky Survey \citep[SDSS,][]{sdss}, positional and 
  photometric data of giant stars with relatively precise distances, 
  including stars on the horizontal branch (HB), were used to map the 
  streams. It was established that the apocentres of the leading and 
  trailing arm are significantly different
  \citep[$47.8\pm0.5$ and $102.5\pm2.5$\,kpc, respectively,][]%
  {belokurov_et_al2014}.
  \citet{belokurov_et_al2014} furthermore reported i) a confirmation of 
  the bifurcation of the southern trailing tail 
  \citep{koposov_et_al2012} from the double-peaked profile of blue HB 
  (BHB) stars across the stream at longitude
  $145\degr \!<\! \Lambda_\mathrm{Sgr} \!<\! 170\degr$ in stream 
  coordinates%
  \footnote{See appendix of \citet{belokurov_et_al2014} for this 
            coordinate transformation},
  (ii) a gentle precession of the orbital plane, and (iii) a 
  potentially differing evolution across the sky of the brighter and 
  fainter tails.
  
  Further insight in the 3D structure of the Sagittarius debris, out to 
  large distances, can be provided by variable RR Lyrae stars
  \citep[e.\,g.,][]{vivas_et_al2005}. These tracers can be uncovered in
  time-domain photometry, as for instance provided by the Panoramic
  Survey Telescope and Rapid Response System 1
  \citep[Pan-STARRS 1, hereafter PS1,][]{ps1}, or currently with much 
  larger coverage and precison by \textit{Gaia} DR3
  \citep{gdr3, clementini_et_al2023, xin_yi_li_et_al2023, 
         muraveva_et_al_arxiv2407_05815}.
  The PS1 dataset resulted in the remarkably clean and complete 
  catalogue of RR Lyrae stars by \citet{sesar_et_al2017a} and 
  \citet{hernitschek_et_al2017}. \citet{sesar_et_al2017b} use this 
  sample within 13\degr\ of the Sagittarius orbital plane to study the 
  Sagittarius debris, resulting in a selection of 19,000 candidate RR 
  Lyrae stars. With their catalogue sensitivity out to larger 
  distances, they reveal an extra arm at heliocentric distances of 
  120\,kpc in the trailing stream at
  $\Lambda_\mathrm{Sgr} \!=\! 172\degr$ called the ``spur'' feature. 
  \citet{starkenburg_et_al2019} subsequently note that their sample of 
  photometrically selected candidate BHB stars also shows the same 
  feature, confirming that indeed the most distant features of the 
  Sagittarius stream are equally complex, with many intricate features. 

  Many studies have demonstrated they can model the multiple wraps of 
  the Sagittarius debris
  \citep[e.\,g.,][]{ibata_lewis1998, helmi_white2001, lm10,
                    dierickx_loeb2017, thomas_et_al2017,
                    fardal_et_al2019, vasiliev_et_al2021,
                    oria_et_al2022, davies_et_al2024}.
  However, only a few studies have made predictions for the spur
  feature under study. For instance, both \citet{dierickx_loeb2017} and 
  \citet{fardal_et_al2019} show that using N-body simulations one can 
  reproduce the multiple wraps as well as the spur feature. As the data
  are consistent with a wide range of models, the authors mention
  several tests for their models, including the velocity structure of
  the spur feature. \citet{fardal_et_al2019} provide a number of
  important model predictions to separate stars dynamically in the spur
  feature from the main trailing arm. These authors predict that line-%
  of-sight (LOS) velocity measurements for a sample of bright spur
  feature members reaching an uncertainty of about
  $10 \, \mathrm{km\,s^{-1}}$ would be sufficient for this purpose.
  
  An additional study to predict the stream feature, and furthermore 
  also model the influence of the LMC on the merger as first suggested
  by \citet{vera-ciro_helmi2013}, is presented by 
  \citet{vasiliev_et_al2021}. Their N-body simulations of the 
  interaction of the LMC, Galaxy, and Sagittarius produce a spur 
  feature composed of stars stripped about 2.5 billion years ago at the 
  penultimate pericentre passage of the Sagittarius galaxy. 
  Subsequently, \citet{oria_et_al2022} re-explore the idea that the 
  progenitor of Sagittarius was a disc galaxy, with the disc roughly 
  perpendicular to both the MW disc and Sagittarius dwarf galaxy orbital
  plane. Their model results in the observed bifurcation of both the
  leading and trailing part of the Sagittarius tidal arm
  \cite[as observed in more detail by, e.\,g.,][]{ramos_et_al2022}, but
  it seems not to significantly affect the Sagittarius spur region.

  In this work, we use updated photometric catalogues and selection 
  techniques \citep{thomas_et_al2018, starkenburg_et_al2019} to select 
  the best candidate BHB member stars of the outermost trailing arm 
  apocenter and spur feature and follow these stars up 
  spectroscopically to determine their LOS velocities and constrain 
  this feature kinematically for the first time. The most luminous
  standard candles observable with sufficient signal-to-noise ratio are
  outside of the limits of current spectroscopic surveys. Therefore,
  there is little published data on LOS velocities of bright stars in
  this outer realm of the Galaxy. 
    
  This work has two primary aims: firstly, we aim to investigate, using 
  spectroscopic data, the effectiveness of selecting BHB stars using 
  narrow and/or broad-band photometry, especially at the faint-end 
  magnitudes, corresponding to BHB stars located in the outermost region
  of the MW stellar halo at distances above 50\,kpc. To this end, we 
  employ the ESO/VLT/FORS2 spectrograph with the 600B+22 grating 
  covering the Balmer lines of the stars from which their 
  discriminatory surface gravity can be determined, separating BHB 
  stars from contaminants.%
  \footnote{Based on observations collected at the European Southern 
            Observatory under ESO programmes 106.21L8.001 and
            106.21L8.002 (PI: E. Starkenburg).}
  Secondly, we wish to ascertain whether the LOS velocities derived from
  the optical VLT/FORS2 spectra of the confirmed BHB stars can provide
  further observational constraints on the MW gravitational potential at
  large distance.

  The structure of this study is as follows. In \autoref{sec:data} 
  we lay out the characteristics of the sample of candidate BHB 
  stars in the Sagittarius stream spur feature, describe the
  methodology used for the data reduction, spectral analysis, and
  stellar classification, and present the data on the kinematics of the 
  classified BHB stars. We present and discuss the results in comparison
  to simulations and its implications for the Galactic potential at
  large distances (\autoref{subsec:mwprofile}) and to the MW globular
  cluster NGC\,2419 (\autoref{subsec:n2419}) in \autoref{sec:results},
  and summarise in \autoref{sec:sum}.

  \section{Data}\label{sec:data}
  
  \subsection{Sample selection of candidate Blue Horizontal Branch    
              stars}

  In order to cleanly select BHB stars, one needs to discriminate them
  from contamination sources with similar broad-band colours, such as
  white dwarfs, quasars, and -- most importantly at the magnitudes of
  our study -- blue straggler stars. Previous research has
  established that there are various ways to deselect spectroscopically 
  and/or photometrically the mentioned contaminants
  \citep[e.\,g.,][]{xue_et_al2008, xue_et_al2011, vickers_et_al2012, 
                    barbosa_et_al2022, fengqing_yu_et_al2024, 
                    amarante_et_al2024, tetsuya_fukushima_et_al2024, 
                    bystroem_et_al_arxiv2410_09149}.
  
  Regarding quasars as contamination, analysis of the dereddened
  $(u\!-\!g) _ {\text{SDSS},0}$ versus $(g\!-\!r) _ {\text{SDSS},0}$
  diagram of stellar-like objects by \citet{yanny_et_al2000} showed that
  A-type stars have similar $(g\!-\!r) _ {\text{SDSS},0}$ colour index
  values as quasars but are about 1\,mag redder in 
  $(u\!-\!g) _ {\text{SDSS},0}$. Using this or a similar colour space, 
  previous works have been able to select out quasars as well as white 
  dwarfs 
  \citep[e.\,g.,][]{xue_et_al2008, xue_et_al2011, deason_et_al2011, 
                    deason_et_al2014, ibata_et_al2017, 
                    starkenburg_et_al2019, fantin_et_al2019, 
                    barbosa_et_al2022}.
                        
  Other works have looked at colour spaces using the near-infra-red
  instead of the near-ultraviolet in combination with optical photometry
  and report also success in filtering out quasars and white dwarfs 
  from (candidate) A-type stars
  \citep[e.\,g.,][]{vickers_et_al2012, deason_et_al2018,
                    tetsuya_fukushima_et_al2018,
                    tetsuya_fukushima_et_al2019, fengqing_yu_et_al2024, 
                    amarante_et_al2024, tetsuya_fukushima_et_al2024}.
  A key study of \citet{vickers_et_al2012} comparing SDSS
  $(i\!-\!z) _ {0,\text{SDSS}}$ and $(g\!-\!r) _ {0,\text{SDSS}}$ of
  stellar-like objects with spectroscopic data found that quasars and
  white dwarfs are on average the reddest and bluest, respectively, in
  $(i\!-\!z) _ {0,\text{SDSS}}$ of all contaminants.To further
  distinguish BHB stars from quasars, \citet{vickers_et_al2012}, in turn
  based on the results of \citet{lenz_et_al1998}, also use
  $(g\!-\!r) _ {\text{SDSS},0}$, since quasars tend to have higher
  values for this colour index (especially at higher redshifts) due to
  more uniform emission in these bands, in contrast to blue stars. Most
  studies investigating the photometric selection of (candidate) BHB
  stars since have indeed been based on such a combination of optical
  and near-infra-red filters using either data from the Subaru
  Telescope/Hyper Suprime-Cam
  \citep[e.\,g.,][]{deason_et_al2018, tetsuya_fukushima_et_al2018,
                    tetsuya_fukushima_et_al2019,
                    tetsuya_fukushima_et_al2024}
  or the Dark Energy Survey \citep[e.\,g.,][]{fengqing_yu_et_al2024}.
  
  To tackle the contamination from blue straggler stars at similar
  magnitudes, mainly three ways are suggested in the literature.
  Firstly, some studies have used typically low-resolution optical
  spectra of A-type stars to use the measured widths of Balmer lines as
  discriminator between BHB and more compact blue straggler stars. Due
  to their higher surface gravity and thus stronger pressure broadening,
  the Balmer lines are broader for blue straggler stars than for BHB
  stars
  \citep[see, e.\,g.,][]{clewley_et_al2002, xue_et_al2008,
                         xue_et_al2011}.
  A discrimination based on the widths of Balmer lines can be expressed 
  in these measured widths directly, but also in resulting effective 
  temperatures and surface gravities estimated from (these same lines 
  in the) spectra
  \citep[e.\,g.,][]{barbosa_et_al2022, bystroem_et_al_arxiv2410_09149}.
  \citet{vickers_et_al2021} demonstrate that also a classifier trained
  on Large Sky Area Multi-Object Fiber Spectroscopic Telescope
  \citep[LAMOST,][]{lamost} spectra of BHB and non-BHB stars with
  sufficient signal-to-noise ratio can yield a selection of (candidate)
  BHB stars with a purity of 86\,per\,cent.
  
  Secondly, as suggested by \citet{lenz_et_al1998} a near-ultraviolet 
  band as SDSS \textit{u} that captures the Balmer break in hot stars 
  can also be used to differentiate between BHB and blue straggler 
  stars since the strength of the Balmer break in hot stars is 
  dependent on the surface gravity
  \citep[see works by, e.\,g.,][]%
  {ruhland_et_al2011, deason_et_al2011, deason_et_al2014,
   thomas_et_al2018, starkenburg_et_al2019}.
   
  Thirdly, it is now well established from a variety of studies that 
  also near-infra-red filters around the Paschen lines in combination 
  with optical photometry also separates blue straggler from BHB stars
  when the signal-to-noise ratio is high enough
  \citep[e.\,g.,][]{vickers_et_al2012, deason_et_al2018, 
                    tetsuya_fukushima_et_al2018,
                    tetsuya_fukushima_et_al2019, thomas_et_al2018, 
                    fengqing_yu_et_al2024, amarante_et_al2024, 
                    tetsuya_fukushima_et_al2024}. 
                        
  In summary, it has been shown that even photometric selections of 
  candidate BHB stars can provide high purities between 80 and 90 per
  cent. The approach we follow in this work for photometric selection 
  of candidate BHB stars in the spur feature of the Sagittarius stream 
  uses this technique as well.
  
  Following the approach of \citet{thomas_et_al2018}, we initially 
  select 974 candidate BHB stars in the Galactic halo. These candidate
  BHB stars were originally isolated from contaminating candidate blue
  straggler stars in a photometric sample that combined PS1 $griz$ data
  taken as part of the 3$\Pi$ survey with \textit{u}-band data taken
  with CFHT/MegaCam. The latter observations were taken as part of the
  Canada-France Imaging Survey \citep[CFIS,][]{ibata_et_al2017}, that
  has since been absorbed into the Ultraviolet Near Infrared Optical
  Northern Survey (UNIONS; Gwyn et al. 2025, in prep). In addition to
  CFIS, UNIONS consists of members of the Pan-STARRS team, and the Wide
  Imaging with Subaru HyperSuprime-Cam of the \textit{Euclid} Sky
  (WISHES) team. CFHT/CFIS is obtaining deep \textit{u} and \textit{r}
  bands; Pan-STARRS is obtaining deep \textit{i} and moderate-deep
  \textit{z} band imaging, and Subaru is obtaining deep \textit{z}-band
  imaging through WISHES and \textit{g}-band imaging through the
  Waterloo-Hawaii IfA \textit{g}-band Survey (WHIGS). These independent
  efforts are directed, in part, to securing optical imaging to
  complement the \textit{Euclid} space mission \citep{euclid}, although
  UNIONS is a separate collaboration aimed at maximizing the science
  return of these large and deep surveys of the northern skies. We note
  that with the advent of these deeper bands using other facilities, the
  analysis of \citet{thomas_et_al2018} could be redone using deeper
  data. However, here we use the existing BHB catalog and demonstrate it
  is more than sufficient to meet our science goals.
  
\begin{figure*}
    \centering
  	\includegraphics{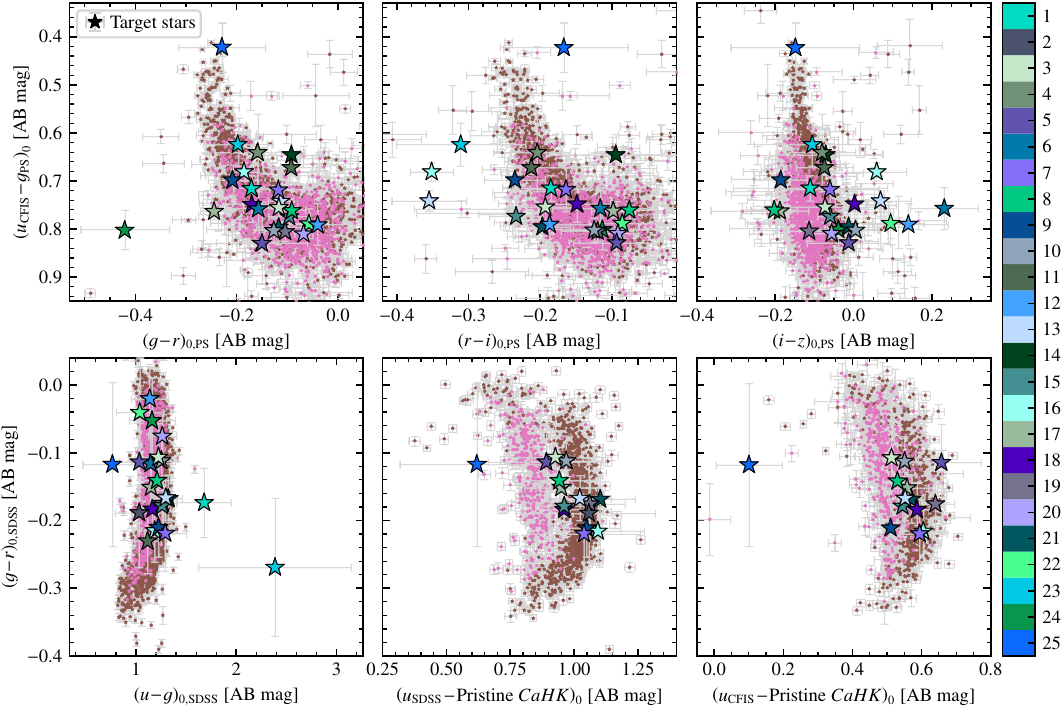}
  	\caption{An overview of the target stars in multiple colour-colour 
  	         spaces is shown. Data were gathered from multiple sources. 
             These are: the Panoramic Survey Telescope and Rapid
             Response System 1 3$\Pi$ survey Data Release 2
             \citep{ps1, magnier_et_al2020}, 
             the Canada-France Imaging Survey, the Sloan Digital Sky
             Survey Data Release 18 \citep{sdss_dr18}, and
             the Pristine Survey \citep{starkenburg_et_al2017}. The 
             candidate BHB stars in our sample are colour-coded
             according to the number of their id given in 
             \autoref{tab:cand_bhb_stars}. The background stars come 
             from two samples. The pink markers are blue straggler 
             stars from the \citet{xue_et_al2008} set while the brown
             symbols represent BHB stars from the \citet{xue_et_al2011} 
             dataset.}
    \label{fig:target_selection}
  \end{figure*}

  The colour spaces, and the location of spectroscopically confirmed
  BHB and blue straggler stars
  \citep[taken from the studies of][]{xue_et_al2008, xue_et_al2011} are
  shown in the upper panels of \autoref{fig:target_selection}. From the
  original 974 stars, 29 candidate BHB stars were identified to be 
  possibly associated to the spur feature due to their proximity in 
  terms of declination (25--40\,deg) and distance ($\sim$80--150\,kpc)
  to this distant arm of the Sagittarius tidal stream. An additional 40
  BHB candidates could be added in the same region from the sample of
  \citet{starkenburg_et_al2019}. While \citet{thomas_et_al2018} focuses
  on the photometric selection of candidate BHB stars based on 
  CFIS-$u_0$ and PS1 $g_0r_0i_0z_0$, \citet{starkenburg_et_al2019} 
  demonstrated the efficacy of employing a combination of broad-band 
  SDSS $ugr_0$ and narrow-band Pristine Survey $CaHK_0$ photometry
  \citep{starkenburg_et_al2017} to identify candidate BHB stars in the
  Galactic halo and isolate them from contaminating samples. The high
  efficiency of the combination of the two blue \textit{u} and
  \textit{CaHK} bands is illustrated in the lower left and middle panel 
  of \autoref{fig:target_selection}. Taken together, the \textit{u} and
  \textit{CaHK} band demarcate the Balmer break and are therefore 
  highly effective to measure the higher surface gravity of
  blue straggler compared to BHB stars.
  
  Figure \ref{fig:target_selection} illustrates the power of the 
  combined colour selection from photometry of DR2 of the PS1 3$\Pi$ 
  survey \citep{ps1}, CFIS \citep{ibata_et_al2017}, SDSS DR18 
  \citep{sdss_dr18}, and the Pristine Survey 
  \citep{starkenburg_et_al2017}. For the estimation of interstellar 
  extinction we used the integrated colour excess values along the 
  LOS of each star from \citet{sfd} in combination with the tabulated
  values
  $A_X / E(B\!-\!V)_\text{SFD}$ from \citet{sf11} assuming
  $R_V \!=\! 3.1$ and
  $A_{\mathrm{Pristine}\ CaHK} / E(B\!-\!V)_\text{SFD} \!=\! 3.918$ 
  \citep{pristine_dr1}, where we assume that extinction in 
  CFIS-\textit{u} is roughly the same as in Sloan-\textit{u} following
  \citet{ibata_et_al2017},
  \citet{thomas_et_al2018, thomas_et_al2019, thomas_et_al2020}, and
  \citet{jensen_et_al2021}.

   \begin{table*}
    \centering
    \caption{Basic properties of the candidate BHB stars in the sample.}
    \begin{tabular}{llllllll}
      \hline
      Designation & Id & R.A. & Decl. & \textit{Gaia} source id
      & Observation time & Source \\
      &  & J2000 & J2000 & & Julian Date-UTC &  \\
      \hline
      \multirow{2}{*}{144511105510814456} & \multirow{2}{*}{candBHB1}
      & \multirow{2}{*}{7h22m12.24s}
      & \multirow{2}{*}{$30^\circ25{}^\prime40.8{}^{\prime\prime}$}
      & \multirow{2}{*}{885923969299049856} & 2459145.820359
      & \multirow{2}{*}{C} \\
      &  & & & & 2459193.744190 & \\
      J072805.06+254349.2 & candBHB2 & 7h28m05.04s
      & $25^\circ43{}^\prime48{}^{\prime\prime}$ & ... & 2459200.699201
      & P \\
      141301122397717196 & candBHB3 & 7h28m57.6s
      & $27^\circ45{}^\prime21.6{}^{\prime\prime}$ & 873208357842184448
      & 2459177.726597 & C \\
      147381140874742260 & candBHB4 & 7h36m20.88s
      & $32^\circ49{}^\prime04.8{}^{\prime\prime}$ & ...
      & 2459193.787593 & C \\
      J074315.57+331906.5 & candBHB5 & 7h43m15.6s
      & $33^\circ19{}^\prime04.8{}^{\prime\prime}$ & ...
      & 2459196.707963 & P \\
      J075310.18+271300.1 & candBHB6 & 7h53m10.08s
      & $27^\circ13{}^\prime01.2{}^{\prime\prime}$ & ...
      & 2459202.715949 & P \\
      148711185503191921 & candBHB7 & 7h54m12s
      & $33^\circ55{}^\prime33.6{}^{\prime\prime}$ & ...
      & 2459197.740104 & C \\
      J075724.50+330505.6 & candBHB8 & 7h57m24.48s
      & $33^\circ05{}^\prime06{}^{\prime\prime}$ & 881458199742504832
      & 2459225.660648 & P \\
      J075815.87+364537.2 & candBHB9 & 7h58m15.84s
      & $36^\circ45{}^\prime36{}^{\prime\prime}$ & 906891484241931264
      & 2459196.753993 & P \\
      J075922.63+265611.1 & candBHB10 & 7h59m22.56s
      & $26^\circ56{}^\prime09.6{}^{\prime\prime}$ & 874378478731783424
      & 2459192.777477 & P \\
      J080051.99+361303.2 & candBHB11 & 8h00m52.08s
      & $36^\circ13{}^\prime04.8{}^{\prime\prime}$ & 906624234198306816
      & 2459200.766829 & P \\
      J080146.11+343120.7 & candBHB12 & 8h01m46.08s
      & $34^\circ31{}^\prime19.2{}^{\prime\prime}$ & 905620380080032768
      & 2459199.732419 & P \\
      J080152.47+253951.0 & candBHB13 & 8h01m52.56s
      & $25^\circ39{}^\prime50.4{}^{\prime\prime}$ & 682026890911273088
      & 2459176.742639 & P \\
      145911206860680542 & candBHB14 & 8h02m44.64s
      & $31^\circ35{}^\prime31.2{}^{\prime\prime}$ & ...
      & 2459198.756331 & C \\
      142131209582933555 & candBHB15 & 8h03m49.92s
      & $28^\circ26{}^\prime38.4{}^{\prime\prime}$ & 876085642331737984
      & 2459192.730613 & C \\
      J080408.41+355917.0 & candBHB16 & 8h04m08.4s
      & $35^\circ59{}^\prime16.8{}^{\prime\prime}$ & 906925603461957248
      & 2459202.755405 & P \\
      J080712.26+300447.1 & candBHB17 & 8h07m12.24s
      & $30^\circ04{}^\prime48{}^{\prime\prime}$ & 876855197392497024
      & 2459201.713171 & P \\
      J080757.78+350947.9 & candBHB18 & 8h07m57.84s
      & $35^\circ09{}^\prime46.8{}^{\prime\prime}$ & 905889932227653888
      & 2459224.701400 & P \\
      J080827.57+301532.4 & candBHB19 & 8h08m27.6s
      & $30^\circ15{}^\prime32.4{}^{\prime\prime}$ & 876954290877415424
      & 2459203.723565 & P \\
      J080943.83+300352.5 & candBHB20 & 8h09m43.92s
      & $30^\circ03{}^\prime54{}^{\prime\prime}$ & 876748269886574720
      & 2459207.693669 & P \\
      J081027.60+315155.4 & candBHB21 & 8h10m27.6s
      & $31^\circ51{}^\prime54{}^{\prime\prime}$ & 901885613958625152
      & 2459177.769769 & P \\
      J081507.72+294705.5 & candBHB22 & 8h15m07.68s
      & $29^\circ47{}^\prime06{}^{\prime\prime}$ & ... & 2459199.792697
      & P \\
      142731243370082351 & candBHB23 & 8h17m20.88s
      & $28^\circ56{}^\prime34.8{}^{\prime\prime}$ & ...
      & 2459176.782975
      & C \\
      \multirow{3}{*}{J084934.60+375926.5} & \multirow{3}{*}{candBHB24}
      & \multirow{3}{*}{8h49m34.56s}
      & \multirow{3}{*}{$37^\circ59{}^\prime27.6{}^{\prime\prime}$}
      & \multirow{3}{*}{...} & 2459196.797465 & \multirow{3}{*}{P} \\
      & & & & & 2459197.792465 & \\
      & & & & & 2459195.808785 & \\
      \multirow{2}{*}{J090045.36+371809.1} & \multirow{2}{*}{candBHB25}
      & \multirow{2}{*}{9h00m45.36s}
      & \multirow{2}{*}{$37^\circ18{}^\prime10.8{}^{\prime\prime}$}
      & \multirow{2}{*}{...} & 2459198.799965 & \multirow{2}{*}{P} \\
      & & & & & 2459225.707431 & \\
      \hline
    \end{tabular}
    \tablefoot{It provides the data on equatorial coordinates in the
               International Celestial Reference System from PS1 DR2
               \citep{ps1, magnier_et_al2020}. We use the SDSS and PS1
               designation in the first column if the star came from the
               Pristine Survey (`P' in last column) 
               \citep{starkenburg_et_al2019} or CFIS catalogue (`C' in
               last column) \citep{thomas_et_al2018}, respectively.}
    \label{tab:cand_bhb_stars}
  \end{table*}
  
  Out of these 69 candidate BHB stars, 25 were followed up 
  spectroscopically using FORS2. These were selected on the basis of a
  combination of their proximity to the stream on the sky declination
  (25--40\,deg where the spur feature is likely located), combined with 
  their colour information. For the remainder of the paper, we will 
  refer to them with their internal IDs for this study
  (second column in \autoref{tab:cand_bhb_stars}).
  
  Figure \ref{fig:target_selection} shows the location of the 25 chosen 
  targets in the various colour-colour spaces that were used for their 
  selection on top of the spectroscopically identified samples. Because 
  of the pilot-programme nature of this observational campaign, it was 
  not enforced that the stars were the best BHB-candidates in all 
  possible colour spaces. The programme was deliberately leaving the 
  option open that one of the colour spaces would deliver an inferior 
  classification. Additionally, we note that updates in the PS1, 
  Pristine survey, and CFIS-\textit{u} more recent (internal) data 
  releases are affecting the colours with respect to the original 
  selection. For Pristine the main updates have been a re-calibration 
  of the narrow-band \textit{CaHK} photometry and updates to the data 
  reduction pipeline \citep{pristine_dr1} and for CFIS the changes 
  between the original catalogue used by  \citet{thomas_et_al2018} and 
  the latest data reduction are due to a change of the photometric 
  reduction pipeline and a better estimation of the zero point. The 
  changes in colour for the targets due to these updates are typically 
  small, as can be appreciated from the fact that in 
  \autoref{fig:target_selection} still most targets lie in the regions 
  covered by BHB stars.

  Most strikingly, we see that candBHB25 is very discrepant in the 
  colour-colour diagrams of the lower and upper left panels compared to 
  where (candidate) BHB stars are to be expected in these projected 
  colour spaces. By contrast, we can also see that the same star is 
  significantly closer to the sequence of (candidate) BHB stars in the 
  upper middle panel of \autoref{fig:target_selection}. Moreover, 
  candBHB24 and candBHB5 are outliers in the top left panel, but fall 
  nicely on the sequence of candidate BHB stars in several other 
  panels. Similarly, many more sample stars are off the range of 
  candidate BHB stars in one of the shown colour spaces, whereas follow 
  the expectations in the other projections illustrated in 
  \autoref{fig:target_selection}.
  
  We note that the majority of candidate BHB stars in the sample have 
  very faint PS1 \textit{g} magnitudes for their Data Release 2, 
  going well beyond 20 (up to 21.6\,AB mag). A consequence of this faint
  selection is the limited overlap of the target stars in \textit{Gaia}
  DR 3, due to incompleteness beyond $Gaia\ G \!\simeq\! 21$\,mag
  \citep{gdr3} (see also \autoref{subsec:n2419}).

  \subsection{VLT/FORS2 spectroscopic setup}

  We collected VLT/FORS2 data in the wavelength range 
  3,300--6,210\,\AA\ using the low spectral resolution ($R \! = \! 780$ 
  at 4,627\,\AA\ with a 1\,arcsec that corresponds to a velocity 
  resolution of
  $\sigma \!=\! c\!/\!(2.35\!\times\!R)%
  = 177\,\text{km}\,\text{s}^{-1}$)
  600B+22 grism for the candidate BHB stars listed in
  \autoref{tab:cand_bhb_stars}. The slit width was 1\,arcsec, exposure
  times are close to 2300\,s, and observational epochs are listed in
  \autoref{tab:cand_bhb_stars} (Period 106, PI: E. Starkenburg; ESO
  programmes 106.21L8.001 and 106.21L8.002).
  
  A key ingredient for reliable and accurate LOS velocity 
  determinations is the quality of the wavelength calibration. 
  In the FORS2 spectrograph, the wavelength calibration is based on the
  lines in spectra of arc-lamps taken at daytime and at zenith. 
  However, for spectra of stars with significant zenith distance (such 
  as at high air masses as in our sample), there is a potential for 
  bias in the wavelength calibrations due to instrument flexure under 
  gravity \citep{fors2_user_p106}. Strong levels of instrument flexure 
  can give rise to second order effects in the wavelength solution.
  
  Many researchers have utilised night sky emission lines and/or 
  telluric lines to measure these second order effects in the 
  dispersion solution that are gathered together with the science 
  spectra in the same setup and under the same influence of instrument 
  flexure
  \cite[e.\,g.,][]{southworth_et_al2006, segue, deason_et_al2012, 
                   caffau_et_al2020}. 
  However, in the wavelength range of our study, only a few suitable
  lines are available. \citet{deason_et_al2012} investigated the
  differential impact of these second order effects using strong night
  sky emission lines in FORS2 spectra of 48 candidate BHB stars in a
  similar setup as in this programme. This study found these effects
  lead to uncertainties in the LOS velocities on the level of
  6\,km\,s$^{-1}$. A more recent study by \citet{caffau_et_al2020}
  involved also a similar FORS2 setup as in this programme and reported
  more variation of the LOS velocities of their stars with reference
  values, after a shift to the dispersion axis of their FORS2 spectra
  based on the measured positions of eight unblended night sky emission
  lines. Following this result, we refrain from this approach.
  
  Instead, to establish whether instrument flexure had a significant 
  impact on the wavelength solution derived from arc-lamps taken at 
  daytime and at zenith, we additionally gathered data with the same 
  spectral setup (except a smaller slit of 0.4\,arcsec) for ten 
  standard stars. Each observing block of our observing programme would 
  typically have a BHB candidate target, immediately followed or 
  preceded by the observation of a velocity standard star as close in 
  the sky as possible. The list of velocity standards, and their LOS 
  velocity measurements from literature, are presented in 
  \autoref{tab:vel_sts} and taken from
  \citet{vel_sts, soubiran_et_al2018}. This approach provides us with a
  second avenue to test the accuracy and precision of the measured LOS 
  velocities. 
  
  \begin{table*}
    \centering
    \caption{Key characteristics of the velocity standard stars.}
    \begin{tabular}{llllllll}
      \hline
      Designation & R.A. & Decl. & Spectral type
      & \multicolumn{2}{l}{Barycentric LOS velocity} \\
      & \multicolumn{2}{l}{J2016} & 
      & \multicolumn{2}{l}{$\left[\text{km}\,\text{s}^{-1}\right]$} \\
      \hline
      BD +24 1843 & 8h04m42.70936675s
      & $24^\circ19{}^\prime49.47351216{}^{\prime\prime}$ & G & 25.168
      & $\pm 0.004$ \\
      BD +31 1781 & 8h18m10.42693323s
      & $30^\circ35{}^\prime49.7809604{}^{\prime\prime}$ & K & 13.457
      & $\pm 0.019$ \\
      BD +34 1955 & 9h12m37.57008557s
      & $33^\circ36{}^\prime01.09932489{}^{\prime\prime}$ & K & 1.539
      & $\pm 0.004$ \\
      WDS J07277+2420A & 7h27m39.94289657s
      & $24^\circ20{}^\prime09.97879538{}^{\prime\prime}$ & K & -18.559
      & $\pm 0.208$ \\
      BD +26 1647 & 7h46m58.48254722s
      & $26^\circ01{}^\prime28.79033441{}^{\prime\prime}$ & G & 13.109
      & $\pm 0.008$ \\
      BD +31 1684 & 7h53m33.99666092s
      & $30^\circ35{}^\prime48.89008376{}^{\prime\prime}$ & G & -234.195
      & $\pm 0.013$ \\
      BD +30 1501 & 7h26m19.81197854s
      & $29^\circ58{}^\prime07.98205929{}^{\prime\prime}$ & K & -36.684
      & $\pm 0.044$ \\
      TYC 2461-988-1 & 7h32m44.21719125s
      & $33^\circ50{}^\prime06.07594227{}^{\prime\prime}$ & F & 23.187
      & $\pm 0.020$ \\
      BD +36 1823 & 8h27m18.62972372s
      & $35^\circ49{}^\prime01.65250699{}^{\prime\prime}$ & F & 9.282
      & $\pm 0.043$ \\
      BD +35 1801 & 8h19m27.87619964s
      & $35^\circ01{}^\prime21.76871662{}^{\prime\prime}$ & F & -23.908
      & $\pm 0.013$ \\
      \hline
    \end{tabular}
    \tablefoot{The data are sourced from \citet{gdr3} except the
               barycentric LOS velocities that come from
               \citet{soubiran_et_al2018}. We used the same approach for
               the designations as noted in
               \autoref{tab:cand_bhb_stars}. The equatorial coordinates
               are also in the International Celestial Reference System.
               Spectral types come from the Extended Stellar
               Parametrizer for Hot Stars module as part of
               \textit{Gaia} DR3 Astrophysical parameters inference
               system \citep{creevey_et_al2023}.}
    \label{tab:vel_sts}
  \end{table*}
  
  \subsection{Spectroscopic reduction}

  Standard two-dimensional, CCD data reduction and calibration were 
  performed using the FORS2 pipeline 5.5.7 (2021) within the 
  \textsc{\lowercase{EsoReflex}} 2.11.3 (2021) environment
  \citep{freudling_et_al2013}. Details are given in
  \autoref{appendix_sec:red_calib}%
  \footnote{Data of the first observation of candBHB1 was sourced from 
            the ESO archive later as part of the open stream release 
            FOR S2-SPEC. Based on data obtained from the ESO Science 
            Archive Facility with DOI:
            https://doi.eso.org/10.18727/archive/77.}.
  
\begin{figure}
    \centering
    \includegraphics{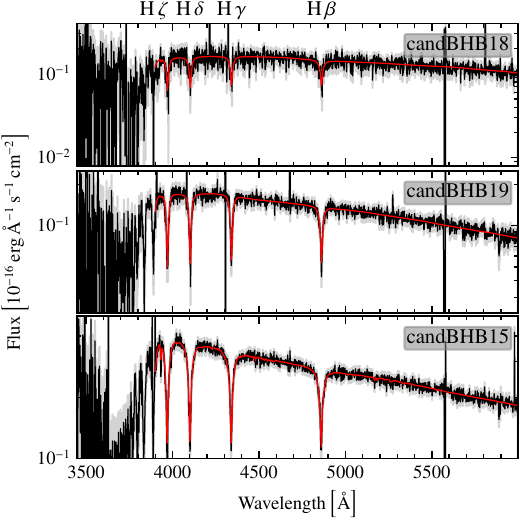}
    \caption{The example spectra of three candidate BHB stars listed in
             \autoref{tab:cand_bhb_stars} are shown where the 
             wavelength scale is relative to the spectrograph's 
             reference frame at the time of each observation. 
             Background shaded areas represent the uncertainties of the 
             spectra. Solid red lines indicate best-fitting models. The 
             labels at the top annotate some of the strong Balmer lines 
             visible in the spectra. We choose to show the spectrum of 
             the star with the lowest, median of, and highest median 
             signal-to-noise ratio of all spectra across the dispersion 
             axis considered for the fitting in the top, middle, bottom 
             panel, respectively.}
    \label{fig:spectra}
  \end{figure}

  The results of the reduction and calibration are set out in 
  \autoref{fig:spectra} for three example spectra of candidate BHB 
  stars in our sample. While most of the targets were observed with one 
  single exposure, our faintest programme targets have multiple 
  exposures. The spectra have signal-to-noise ratios in the range 7-25.
  This figure depicts the general pattern of a prominent blue continuum
  and Balmer lines of most of the shown spectra of the candidate BHB
  stars, as we would expect for these types of stars.%

  The strong and narrow peaks at various wavelengths in the top two 
  shown spectra in \autoref{fig:spectra} are caused by pixels 
  in the CCD data in the area of the spectra affected by cosmic rays.
  While it would have been an option to include a removal of the 
  effects of cosmic rays in the FORS2 pipeline in a fore- and 
  background subtraction based on a global sky spectrum, a global sky 
  subtraction does not take into account variability of the spectral 
  resolution across the slit axis. In addition, given that the cosmic
  rays present in the spectra of some candidate BHB stars do not affect
  all the Balmer lines that will be used to identify true BHB stars and 
  misclassified BHB stars, we choose to keep them in the spectra.  
  
  The calibrated fluxes rise and fall steeply around 5570--5580\,\AA\ 
  for the shown spectra in \autoref{fig:spectra}. This is probably due 
  to the non-optimal performance of the sky subtraction in that 
  wavelength region where the strong [O\,\textsc{\lowercase{I}}] night 
  sky emission line is located. 

  In \autoref{fig:spectra} there is a clear trend of decreasing 
  signal-to-noise noise blueward of around 3,800\AA. This could be 
  attributed to the following factors. Firstly, it is known that the 
  blue flat field lamp is unstable below 3,800\AA\ leading to 
  variations of the spectral energy distribution of 20\,per\,cent and
  systematic distortions in the calibrated flux given that flat fields
  are obtained separately for science and flux standard star
  observations \citep{fors2_user_p106}. Secondly, extrapolation of the
  wavelength solution for the blue end due to a lack of spectral lines
  in the arc-lamps at the blue edge results in significant residuals in
  this part of the spectrum. This does not affect the central regions of
  the spectra where most of the Balmer lines are located. It can
  therefore be assumed that the noise-dominated, blue part of the
  spectrum has no impact on our analysis. In the remainder of this work,
  we considered only 3,900\AA\ to 5,990\AA\ to be the reliable range of
  the spectrum.

  \subsection{Spectral fitting}
  
  We used \textsc{\lowercase{RVSpecFit}}
  \citep{koposov_et_al2011, koposov2019}%
  \footnote{version 0.4.0.240116+dev at github.com/segasai/rvspecfit}, 
  to analyse the spectra. A major advantage of spectral fitting with 
  \textsc{\lowercase{RVSpecFit}} is that it provides both estimates of 
  the LOS velocity and of effective temperature and surface 
  gravity from each spectrum. Whereas the first parameter provides us 
  the data to constrain the velocity structure of the spur feature of 
  the Sagittarius stream, the latter makes it possible to classify our 
  sample of candidate BHB stars into bona fide BHB, blue straggler
  stars, or other contaminants in a similar way as done by
  \citet{barbosa_et_al2022} and \citet{bystroem_et_al_arxiv2410_09149} 
  (\citet{bystroem_et_al_arxiv2410_09149} also uses results from 
  \textsc{\lowercase{RVSpecFit}}).
  
  To estimate the LOS velocity, effective temperature, and
  surface gravity of a star from a spectrum,
  \textsc{\lowercase{RVSpecFit}} determines the best-fitting synthetic
  spectrum from a set of interpolated
  synthetic stellar spectra from the PHOENIX library
  \citep{husser_et_al2013}, version 2. The shift results in a measured 
  apparent LOS velocity of the star. Any remaining imprecision in the 
  data is taken into account by a multiplicative, normalizing 
  polynomial. In more detail, \textsc{\lowercase{RVSpecFit}} maximises 
  a multivariate Gaussian likelihood that describes a stellar spectrum 
  given the effective temperature, $\log_{10}(g)$, [Fe/H], and
  [$\alpha$/Fe] as parameters of PHOENIX library of templates and LOS 
  velocity%
  \footnote{While \textsc{\lowercase{RVSpecFit}} includes the capability
            to estimate the stellar rotation parameter $v\sin\ i$,
            in section 6 about the known limitations of the Dark Energy
            Spectroscopic Instrument Early DR based also on results from
            \textsc{\lowercase{RVSpecFit}}
            \citet{koposov_et_al_arxiv2407_06280} recommend not to use
            estimates $v\sin\ i$ from the current version of
            \textsc{\lowercase{RVSpecFit}} due to several issues as
            mentioned in \citet{koposov_et_al_arxiv2407_06280}.
            Because of this and also the irrelevance of this parameter
            for the scope of this work, we will not consider the output
            $v\sin\ i$ by \textsc{\lowercase{RVSpecFit}}.}
  that results after marginalising over the cofficients of the
  normalising polynomial given in the initial $\chi^2$ of the
  spectroscopic data given a template of the prepared and interpolated
  PHOENIX grid.
  
  To begin the process of spectral fitting with
  \textsc{\lowercase{RVSpecFit}}, we had to prepare the set of synthetic
  stellar spectra from the PHOENIX library, in the three
  instrumental setups that we use and interpolate between the grid
  parameters. The three instrumental setups that we use are connected 
  to the spectra of the candidate BHB stars in our sample taken with 
  the slit with a width of 1.0\,arcsec, the spectra of the velocity 
  standard stars taken with the 0.4\,arcsec slit, and Sloan Digital Sky 
  Survey DR17 spectra \citep{SDSS_DR17} of a set of reference BHB and 
  blue straggler stars that we use to classify our target stars in the
  effective temperature versus surface gravity space. 
  
  The process of parameter estimation was carried out given the observed
  spectra and priors for each parameters. By default,
  \textsc{\lowercase{RVSpecFit}} assumes uniform priors of
  effective temperature, $\log_{10}(g)$, [Fe/H], and
  [$\alpha$/Fe] over the entire PHOENIX grid although Gaussian priors
  can be custom defined. As detailed below, we use this option for
  parameters we have prior information for. We additionally adopt a
  broad uniform prior of the LOS velocity between $-$500\,km\,s$^{-1}$
  and 500\,km\,s$^{-1}$ throughout all performed fitting with 
  \textsc{\lowercase{RVSpecFit}} to make no strong prior assumptions 
  for the most important parameter of interest in this study.
    
  For the spectra of the velocity standard stars,
  \autoref{tab:norm_priors_rvspecfit_pars_vel_sts} lists 
  the definitions of the normally-distributed priors of effective
  temperature, $\log_{10}(g)$, [Fe/H], and [$\alpha$/Fe] that we adopted
  for the posterior calculation given within
  \textsc{\lowercase{RVSpecFit}}. These utilise estimates of effective
  temperature, $\log_{10}(g)$, [Fe/H], and [$\alpha$/Fe] from
  \textit{Gaia} DR3 the \textit{Gaia} General Stellar Parametriser for
  Spectra catalogue \citep{gdr3, recio-blanco_et_al2023} and sample
  statistics for the velocity standard stars for which no estimates of
  [Fe/H] and [$\alpha$/Fe] was available in this catalogue.
  
  \begin{table*}
    \centering
    \caption{Summary of normally-distributed priors of RVSpecFit
             \textsc{\lowercase{RVSpecFit}}
             \citep{koposov_et_al2011, koposov2019} stellar parameters
             used for the inference of the LOS velocities from the
             VLT/FORS2/600B+22 grism spectra taken in the slit with
             $\text{width} \!=\! 0.4$\,arcsec of the velocity standard
             stars with \textsc{\lowercase{RVSpecFit}}.}
    \begin{tabular}{llllllllllllll}
      \hline
      Designation & \multicolumn{2}{l}{$T_\text{eff}$}
      & \multicolumn{2}{l}{$\log_{10}\ g$} & \multicolumn{2}{l}{[Fe/H]}
      & \multicolumn{2}{l}{[$\alpha$/Fe]}
      & \multicolumn{2}{l}{Source [Fe/H]}
      & \multicolumn{2}{l}{Source [$\alpha$/Fe]}\\
      & [K] & [K] & & & & & & & & & &\\
      (1) & (2) & (3) & (4) & (5) & (6) & (7) & (8) & (9) & (10) & (11) 
      & (12) & (13)\\
      \hline
      BD +24 1843 & 5570 & 33 & 4.17 & 0.11 & -0.1 & 0.2 & -0.01 & 0.0 
      & GSPS & GSPS & GSPS & GSPS \\
      BD +31 1781 & 4532 & 36 & 4.37 & 0.05 & -0.1 & 0.3 & 0.03 & 0.0
      & GSPS & GSPS & GSPS & GSPS \\
      BD +34 1955 & 5235 & 36 & 3.69 & 0.07 & -0.4 & 0.3 & 0.04 & 0.0
      & GSPS & GSPS & GSPS & GSPS \\
      WDS J07277+2420A & 4201 & 150 & 4.24 & 0.20 & -0.2 & 1.4 & -0.2
      & 0.1 & med & max & GSPS & GSPS \\
      BD +26 1647 & 5894 & 510 & 4.89 & 2.55 & -0.2 & 1.4 & 0.07 & 0.2 
      & med & max & GSPS & GSPS \\
      BD +31 1684 & 4250 & 1500 & 1.89 & 2.84 & -0.2 & 1.4 & 0.04 & 0.4 
      & med & max & med & max \\
      BD +30 1501 & 4540 & 33 & 2.11 & 0.08 & -0.4 & 0.3 & 0.03 & 0.0
      & GSPS & GSPS & GSPS & GSPS \\
      TYC 2461-988-1 & 6475 & 180 & 4.19 & 0.29 & -0.2 & 1.4 & 0.2
      & 0.4 & GSPS & GSPS & GSPS & GSPS \\
      BD +36 1823 & 6041 & 45 & 3.98 & 0.05 & -0.2 & 0.3 & 0.06 & 0.0
      & GSPS & GSPS & GSPS & GSPS \\
      BD +35 1801 & 5972 & 66 & 4.09 & 0.11 & -0.1 & 0.3 & 0.07 & 0.1
      & GSPS & GSPS & GSPS & GSPS \\
      \hline
    \end{tabular}
    \tablefoot{The mean of the Gaussian priors of the effective
               temperature (Col. (2)) are the estimates from the
               \textit{Gaia} DR3 \textit{Gaia} General Stellar
               Parametriser for Spectra (GSP-Spec) catalogue
               \citep{gdr3, recio-blanco_et_al2023} that includes
               measurements from the \textit{Gaia} Radial Velocity
               Spectrometer \citep{cropper_et_al2018}. In Col. (3) the
               uncertainties of the GSP-Spec estimates of the effective
               temperature were scaled by the factor 3 to generate a
               broad Gaussian that mimics to some degree an uniform
               prior. In the same vein, we apply the same scaling for
               the other columns that give the standard deviations of 
               the Gaussian priors (Cols. (5), (7), and (9)). 
               Similarly, the data in Cols. (4) and (5) were derived 
               from the GSP-Spec catalogue, using the calibration as 
               recommended in \citet{recio-blanco_et_al2023} (equation 
               1 with polynomical coefficients in
               \citealt{recio-blanco_et_al2023}) to correct the bias for
               their estimates. The means and standard deviations of 
               the broad Gaussian priors of [Fe/H] (Cols. (6) and (7)) 
               and [$\alpha$/Fe] (Cols. (8) and (9)) are based on 
               calibrated values from the GSP-Spec catalogue if 
               available following the recommendations in 
               \citet{recio-blanco_et_al2023}, that is, in case of [M/H]
               using equation 2 with polynomial coefficients given in
               second row of table 3 in \citet{recio-blanco_et_al2023}
               to calculate
               $[\text{Fe}/\text{H}]%
               \!=\![\text{Fe}/\text{M}]+[\text{M}/\text{H}]$
               and with respect to [$\alpha$/Fe] using equation 3 with
               coefficients of the fourth order polynomial in table 4 in
               \citet{recio-blanco_et_al2023} (second [$\alpha$/Fe] of 
               table 4). For some of the velocity standard stars, no 
               GSP-Spec estimates of [M/H] and/or [$\alpha$/Fe] were 
               available. For these stars we took the median of all the 
               other means of [Fe/H] and [$\alpha$/Fe] as the mean of
               the Gaussian prior and the maximum of the other standard 
               deviation entries regarding the standard deviation of 
               the Gaussian prior. In the last four columns GSPS is used
               here to refer to estimates from the GSP-Spec catalogue,
               `med' indicates that for the entry of the mean of the
               Gaussian prior of [Fe/H] or [$\alpha$/Fe] we used the
               median approach as explained above, and `max' labels an
               entry where we adopted the maximum standard deviation of
               all Gaussian priors of [Fe/H] or [$\alpha$/Fe] as also
               noted above.}
    \label{tab:norm_priors_rvspecfit_pars_vel_sts}
  \end{table*}

  As shown in \autoref{tab:norm_priors_eff_temp_cand_bhb_stars}, we do
  not set priors on $\log_{10}(g)$ for our programme stars as we measure
  this value to make the BHB/blue straggler distinction. We have
  information on $T _ \text{eff}$ from their colours (see 
  \autoref{appendix_sec:linear_ps1gr0_color_idx_eff_temp_rel}) and 
  adopt a (conservative) prior based on this information. 
  
  \begin{table}
    \centering
    \caption{Definitions of normally-distributed priors of effective
             temperature used for the inference of the LOS velocities
             from the FORS2 spectra taken in the slit with
             $\text{width} \!=\! 1.0$\,arcsec of the candidate blue
             horizontal branch stars with \textsc{\lowercase{RVSpecFit}}
             \citep{koposov_et_al2011, koposov2019}.}
    \begin{tabular}{lll}
      \hline
      Id & $T_\text{eff}$ & $\Delta T_\text{eff}$ \\
      & [K] & [K] \\
      \hline
      candBHB1 & 8423 & 562 \\
      candBHB2 & 8376 & 842 \\
      candBHB3 & 8140 & 594 \\
      candBHB4 & 8361 & 1362 \\
      candBHB5 & 8317 & 1447 \\
      candBHB6 & 8354 & 994 \\
      candBHB7 & 8144 & 854 \\
      candBHB8 & 8002 & 429 \\
      candBHB9 & 8624 & 773 \\
      candBHB10 & 8192 & 463 \\
      candBHB11 & 8009 & 889 \\
      candBHB12 & 7729 & 1129 \\
      candBHB13 & 8119 & 401 \\
      candBHB14 & 8006 & 975 \\
      candBHB15 & 8030 & 655 \\
      candBHB16 & 8502 & 421 \\
      candBHB17 & 8817 & 440 \\
      candBHB18 & 8420 & 856 \\
      candBHB19 & 8072 & 676 \\
      candBHB20 & 7878 & 281 \\
      candBHB21 & 8140 & 704 \\
      candBHB22 & 7821 & 627 \\
      candBHB23 & 8567 & 1097 \\
      candBHB24 & 9756 & 1409 \\
      candBHB25 & 8735 & 1372 \\
      \hline
    \end{tabular}
    \tablefoot{A relation was approximated from set of BHB stars from 
               \citet{barbosa_et_al2022} that have both PS1 DR2 
               \citep{ps1} \textit{gr} photometry and effective
               temperature values from SEGUE
               (\autoref{%
                  appendix_sec:linear_ps1gr0_color_idx_eff_temp_rel}).
               In the third column uncertainties of effective 
               temperatures resulting from propagation of uncertainties 
               in $g _ \text{PS}$ and $r _ \text{PS}$ and the relation
               including quoted residual variance were scaled by the
               factor 3 to generate a broad Gaussian that mimics to some
               degree an uniform prior.}
    \label{tab:norm_priors_eff_temp_cand_bhb_stars}
  \end{table}

  Finally, we set the priors included in the posterior calculation when 
  fitting the SDSS spectra of BHB and (candidate) blue straggler stars
  in the \citet{barbosa_et_al2022} and \citet{xue_et_al2008} 
  catalogues, respectively, with \textsc{\lowercase{RVSpecFit}}. The
  normally-distributed priors used here were synthesised using the same
  method detailed for the spectra of the candidate BHB stars, with
  effective temperature values sourced from two sources, depending on
  whether it is a BHB or (candidate) blue straggler star, from the
  \citet{barbosa_et_al2022} or \citet{xue_et_al2008} sets, respectively.
  These are the results of the Sloan Extension of Galactic Understanding
  and Exploration (SEGUE) Stellar Parameter Pipeline, as presented by
  \citet{young_sun_lee_et_al2008a, young_sun_lee_et_al2008a} and based
  on SEGUE data \citep{segue} for the BHB stars. In the
  \citet{xue_et_al2008} sample, we adopt the effective temperature
  of the model star with the ELODIE template
  \citet{prugniel_soubiran2001} that best describes the SDSS spectrum 
  of each (candidate) blue straggler star. The latter values are stored
  as metadata in the SDSS spectral data. In contrast to the former
  source, the metadata only have the effective temperatures without
  uncertainties. Consequently, we assumed a uniform, conservative
  standard deviation of 1,000 K for all normally-distributed priors of
  effective temperature when fitting SDSS spectra of the (candidate)
  blue straggler stars with \textsc{\lowercase{RVSpecFit}}. 

  Another significant aspect of the spectral fitting with
  \textsc{\lowercase{RVSpecFit}} is the set of radial basis functions
  used to describe the normalising polynomial. A set of ten radial basis
  functions was used, following the procedure described in 
  \citet{andrew_p_cooper_et_al2023} used by the MW survey pipeline of 
  the Dark Energy Spectroscopic Instrument (DESI-MWS).
  
  \textsc{\lowercase{RVSpecFit}} for all our targets 
  outputs a maximum a posteriori estimation of the LOS velocity while 
  also computing higher moments of the projection of the posterior 
  across the velocity axis (including standard deviation). The 
  estimates of the uncertainty of the maximum a posteriori values of
  $T _ \text{eff}$ and $\log_{10}(g)$ show standard deviations of a 
  Gaussian that was used to approximate the four-dimensional posterior 
  of (effective temperature, $\log_{10}(g)$, [Fe/H], [$\alpha$/Fe]) 
  around the mode. This approximation is equivalent to evaluating the 
  Hessian matrix of the posterior at the mode to obtain the covariance 
  matrix of the Gaussian \citep{bailer-jones2017} as done in 
  \textsc{\lowercase{RVSpecFit}}. 

  \subsection{Line-of-sight velocities and assessment of accuracy of 
              their measurement}%
  \label{subsec:accuracy_los_vels}
  
  As our data reduction does not include any heliocentric correction and
  conversion to the Local Standard of Rest, they are in the observer's 
  reference frame. Heliocentric velocity corrections and conversion to 
  the Local Standard of Rest were calculated using 
  \textsc{\lowercase{noao.rv.rvcorrect}} in the Image Reduction and 
  Analysis Facility
  \citep[\textsc{\lowercase{IRAF}},][]{tody1986, tody1993, iraf}. All
  these calculations were carried out using \textsc{\lowercase{PyRAF}},
  that is, a command language for \textsc{\lowercase{IRAF}} based on
  \textsc{\lowercase{Python}}. The corrections were computed using the
  coordinates of the stars in \autoref{tab:cand_bhb_stars} and
  \autoref{tab:vel_sts}, mid-observation times calculated from the data
  in \autoref{tab:cand_bhb_stars}, and total magnitude of the cartesian
  Galactic velocity (11.1, 12.24, 7.25)\,km\,s$^{-1}$ of the Sun with 
  respect to the Local Standard of Rest as found by 
  \citet{schoenrich_et_al2010}. 

\begin{figure}
    \includegraphics{%
      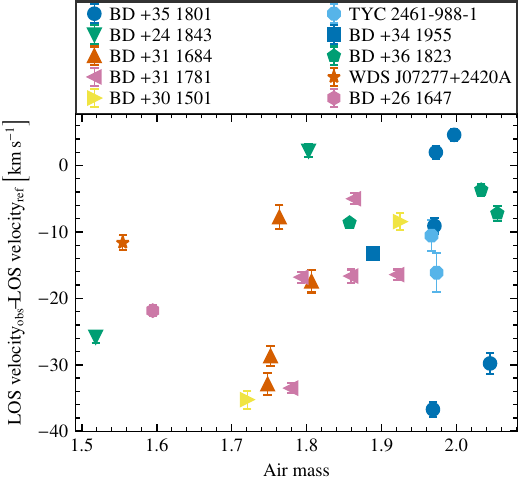}
    \caption{The results obtained from the analysis of the
             FORS2 spectra of the velocity standard stars that were 
             collected from the data of the observations in
             connection with the entries in \autoref{tab:vel_sts}
             compared to the reference barycentric LOS velocity values
             by \citet{soubiran_et_al2018} (also given
             in \autoref{tab:vel_sts}). We plot the estimates of the 
             difference of heliocentric/barycentric LOS velocities
             against the air mass of the star at the time of 
             observation.}
    \label{fig:comparison_los_vel_estimates_vel_sts_main}
  \end{figure}

  The main source of error in our heliocentric LOS velocity estimates is
  a potential for bias resulting from the telescope's exposure to
  flexure under gravity during observation of high air mass target stars
  with the FORS2 instrument. This bias is not accounted for in the
  wavelength solution of the spectra, which is derived based on
  reference spectra of arc lamps obtained with the telescope pointing
  towards zenith. In order to assess whether and how the maximum a
  posteriori estimates of the heliocentric LOS velocities (hereafter
  LOS velocities) are affected, we measured LOS velocities from the
  fitting of the spectra of the velocity standard stars listed in
  \autoref{tab:vel_sts} with \textsc{\lowercase{RVSpecFit}} and compared
  them to the fiducial values in \autoref{tab:vel_sts} from
  \citet{soubiran_et_al2018}. As shown in
  \autoref{fig:comparison_los_vel_estimates_vel_sts_main}, the observed
  LOS velocities of the velocity standard stars here are offset from the
  data given in the catalogue of \citet{soubiran_et_al2018} on average
  by 10--25\,km\,s$^{-1}$. 
  
  Of interest here is the increase in the discrepancy between the 
  estimates of the LOS velocities of the velocity standard stars and 
  the fiducial values with decreasing air mass. This result is somewhat 
  counterintuitive, as the expectation would be that with rising air
  mass stronger possible second order residuals would affect the 
  wavelength calibration of the spectra more as a result of variations 
  caused by telescope flexure under gravity. 
  
  A possible explanation for the behaviour in 
  \autoref{fig:comparison_los_vel_estimates_vel_sts_main} described
  above is that it is a linear trend superimposed on a systematic offset
  of the order of 10--25 km\,s$^{-1}$ (mean and standard deviation of
  $-$15.6 km\,s$^{-1}$ and 11.7 km\,s$^{-1}$, respectively). A 10--25 
  km\,s$^{-1}$ offset in this study corroborates earlier findings by 
  \citet{caffau_et_al2020}. \citet{caffau_et_al2020} found that as 
  measured LOS velocities for four stars from data in a similar 
  instrumental setup as in our work are compared to reference 
  (fiducial) values they are on average offset by 17 km\,s$^{-1}$ with 
  a $\text{standard\ deviation\ } \!=\! 15$\,km\,s$^{-1}$. However, 
  other studies suggested that such an effect could be also lower
  \citep[around 6\,km\,s$^{-1}$,][]{deason_et_al2012}. 
  
  Quantification of a trend with air mass must be approached with some 
  caution because the strength of the variations of the dispersion axis 
  of the spectra against some reference depends also on how the 
  telescope was positioned before the observations of the velocity 
  standard stars, 
  that is, how much zenith distance it moved when targeting the stars. 
  It is beyond the scope of this paper to examine in detail the effect
  of telescope motion across the sky (between observations) on our
  velocity measurements.  
  
  We investigated the relation between our derived LOS velocity and the
  reference value with observing time and S/N but found no clear trend
  (see \autoref{appendix_sec:los_vels_vel_sts}).
  Because a velocity standard is observed for each programme star, we 
  use for our results both the uncorrected LOS velocity as well as 
  corrected LOS velocity by the offset of the velocity standard closest 
  in time to each BHB star.
  
  \subsection{Classification of candidate Blue Horizontal Branch stars 
              in sample}%
  \label{subsec:class}

  We classify the candidate BHB stars in our sample into bona fida BHB
  stars and other contaminants by comparing the positions of the target
  stars in effective temperature versus $\log_{10}(g)$. To establish the
  relevant parameter space we reference fitting results of SDSS DR17
  spectra of a set of BHB and (candidate) blue straggler stars from the
  \citet{barbosa_et_al2022} and \citet{xue_et_al2008} catalogues through
  the same method  (\textsc{\lowercase{RVSpecFit}}). A major advantage
  of using effective temperature and surface gravity inferred from
  spectral fitting in contrast to utilising the Balmer line widths
  \citep[see, e.\,g.,][]{clewley_et_al2002, xue_et_al2008,
                         xue_et_al2011}
  is that it encapsulates all the information from the spectrum instead
  of only individual lines. \citet{barbosa_et_al2022} and 
  \citet{bystroem_et_al_arxiv2410_09149} identified BHB stars in SEGUE 
  \citep{segue} and DESI \citep{desi}, respectively, in a similar way.

\begin{figure}
  	\includegraphics{%
      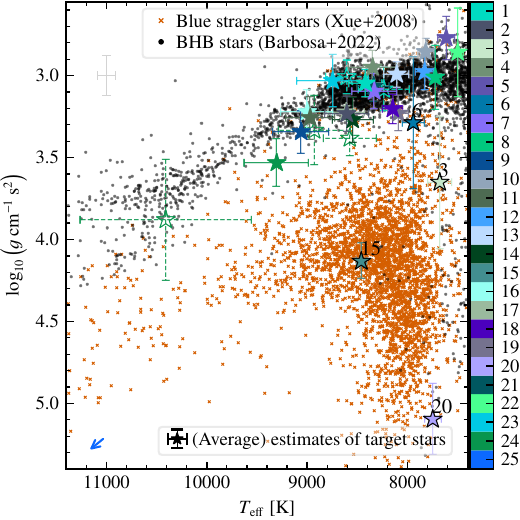}
  	\caption{The trends identified in one projection of the stellar
             parameter space for the group of target stars in the spur
             feature of the Sagittarius stream with FORS2 spectra. A 
             large majority of the estimations of effective temperature 
             and surface gravity for the target stars from 
             \textsc{\lowercase{RVSpecFit}}
             \citep{koposov_et_al2011, koposov2019} (open star markers
             show estimates from fitting spectra obtained from each
             exposure for candBHB1 and candBHB24 for which several
             exposures exist) are in the range expected from bona fide 
             BHB stars (see circles). A few outliers are marked with 
             the numbers and match better the cross symbols that 
             represent a reference dataset of blue straggler stars in 
             \citealt{xue_et_al2008} sample.
             Another outlier, candBHB25, is a likely white dwarf
             candidate highlighted just with an arrow because the best-%
             fitting $T_\text{eff}$ and $\log_{10} g$ for both spectra
             are close to the upper edge of the PHOENIX grid, and are
             therefore not reliable. The grey marker in the upper left 
             corner provides an estimate of the average uncertainties 
             of the reference dataset.}
    \label{fig:class_candidate_bhb_stars_eso_p106programs}
  \end{figure}

  Figure \ref{fig:class_candidate_bhb_stars_eso_p106programs} shows 
  reference results versus our sample of target stars in a Kiel diagram.
  We highlight the low number of contaminants in our sample. Of the
  study population, 21 target stars have (exposure-averaged, maximum a
  posteriori) estimates of effective temperature and $\log_{10}(g)$ 
  consistent with the trend in
  \autoref{fig:class_candidate_bhb_stars_eso_p106programs} of 
  spectroscopically classified BHB stars. The results indicate that 
  candBHB6 could be also in the area of the Kiel diagram where we 
  expect blue straggler stars given its relatively large uncertainties.

  \autoref{tab:results_cand_bhb_stars_eso_p106prgs} shows the proportion
  of different categories of stellar classification of the target stars 
  according to their positions in the two-dimensional space spanned by
  $T _ \text{eff}$ and $\log_{10}(g)$. The bona fide BHB stars are 
  indicated by `BHB' in the `Class' column in 
  \autoref{tab:results_cand_bhb_stars_eso_p106prgs}. The contaminants 
  in our sample classified as blue straggler stars (candBHB15,
  candBHB20, and candBHB3 although with less certainty due to proximity
  to the sequence of BHB stars in
  \autoref{fig:class_candidate_bhb_stars_eso_p106programs}) are marked 
  with `Blue straggler' in the last column in
  \autoref{tab:results_cand_bhb_stars_eso_p106prgs}. One candidate is
  probably a white dwarf (candBHB25). It has unreliable 
  \textsc{\lowercase{RVSpecFit}} output from the spectra gathered from
  the CCD data of two exposures at different dates because they are
  close to the upper limit of the PHOENIX grid (labeled with `WD' and 
  represented with an arrow in
  \autoref{fig:class_candidate_bhb_stars_eso_p106programs}). Less than 
  a third of those candidate BHB stars from the 
  \citet{thomas_et_al2018} catalogue (28\,per\,cent or two out of the 
  seven target stars listed in \autoref{tab:cand_bhb_stars} with PS1 
  ids) are contaminants. This is slightly larger than their estimated 
  contamination fraction of 24\,per\,cent, but this can be attributed 
  to the low-number statistics of our sample. For the subset of 18 
  candidate BHB stars in our sample selected from the 
  \citet{starkenburg_et_al2019} catalogue, only 11.1\,per\,cent are 
  flagged as contamination.

  \begin{table*}
    \centering
    \caption{Spectroscopic parameters (effective temperature,
             $\log_{10}(g)$, LOS velocity) of candidate BHB stars
             obtained with \textsc{\lowercase{RVSpecFit}}.}
    \begin{tabular}{lllllllll}
      \hline
      Id & Time of observation & \multicolumn{2}{l}{MAP $T_\text{eff}$}
      & \multicolumn{2}{l}{MAP $\log_{10}(g)$}
      & \multicolumn{2}{l}{MAP heliocentric LOS velocity}
      & Class \\
      & Julian Date-UTC & \multicolumn{2}{l}{[K]} & &
      & \multicolumn{2}{l}{$\left[\text{km}\,\text{s}^{-1}\right]$} & \\
      \hline
      \multirow{2}{*}{candBHB1} & 2459145.83062068 & 8245 & $\pm\!$253
      & 3.080 & $\pm\!$0.132 & 35.2 & $\pm\!$11.5 & BHB \\
      & 2459193.75057956 & 8611 & $\pm\!$213 & 3.005 & $\pm\!$0.104
      & 9.5 & $\pm\!$13.6 & BHB \\
      candBHB2 & 2459200.70756284 & 8611 & $\pm\!$219 & 3.229
      & $\pm\!$0.108 & $35.4$ & $\pm\!$11.2 & BHB \\
      candBHB3 & 2459177.73637395 & 7691 & $\pm\!$121 & 3.649
      & $\pm\!$0.405 & $78.1$ & $\pm\!$19.8 & Blue straggler \\
      candBHB4 & 2459193.79570399 & 8361 & $\pm\!$238 & 2.947
      & $\pm\!$0.104 & $-30.8$ & $\pm\!$13.6 & BHB \\
      candBHB5 & 2459196.71811152 & 7628 & $\pm\!$85 & 2.766
      & $\pm\!$0.133 & $50.4$ & $\pm\!$12.3 & BHB \\
      candBHB6 & 2459202.72217166 & 7957 & $\pm\!$110 & 3.284
      & $\pm\!$0.403 & $54.4$ & $\pm\!$16.6 & BHB? \\
      candBHB7 & 2459197.75325985 & 8341 & $\pm\!$221 & 3.091
      & $\pm\!$0.109 & $38.9$ & $\pm\!$12.4 & BHB \\
      candBHB8 & 2459225.66818141 & 7737 & $\pm\!$59 & 3.007
      & $\pm\!$0.195 & $27.3$ & $\pm\!$9.0 & BHB \\
      candBHB9 & 2459196.76229864 & 9068 & $\pm\!$278 & 3.340
      & $\pm\!$0.129 & $122.3$ & $\pm\!$8.6 & BHB \\
      candBHB10 & 2459192.7836271 & 7828 & $\pm\!$82 & 2.852
      & $\pm\!$0.083 & $136.3$ & $\pm\!$8.1 & BHB \\
      candBHB11 & 2459200.77374853 & 8975 & $\pm\!$240 & 3.255
      & $\pm\!$0.128 & $31.2$ & $\pm\!$8.7 & BHB \\
      candBHB12 & 2459199.74059709 & 7843 & $\pm\!$92 & 2.973
      & $\pm\!$0.112 & $36.7$ & $\pm\!$11.7 & BHB \\
      candBHB13 & 2459176.74878673 & 8121 & $\pm\!$121 & 2.990
      & $\pm\!$0.077 & $8.9$ & $\pm\!$8.9 & BHB \\
      candBHB14 & 2459198.76587412 & 8553 & $\pm\!$206 & 3.267
      & $\pm\!$0.136 & $71.3$ & $\pm\!$12.4 & BHB \\
      candBHB15 & 2459192.73447173 & 8467 & $\pm\!$52 & 4.125
      & $\pm\!$0.109 & $-48.3$ & $\pm\!$7.9 & Blue straggler \\
      candBHB16 & 2459202.76099233 & 9007 & $\pm\!$256 & 3.216
      & $\pm\!$0.134 & $42.7$ & $\pm\!$12.0 & BHB \\
      candBHB17 & 2459201.72140548 & 8616 & $\pm\!$180 & 3.221
      & $\pm\!$0.042 & $45.0$ & $\pm\!$13.2 & BHB \\
      candBHB18 & 2459224.71022079 & 8158 & $\pm\!$98 & 3.200
      & $\pm\!$0.089 & $87.1$ & $\pm\!$8.2 & BHB \\
      candBHB19 & 2459203.73253058 & 8104 & $\pm\!$99 & 3.226
      & $\pm\!$0.108 & $16.4$ & $\pm\!$9.8 & BHB \\
      candBHB20 & 2459207.70610055 & 7760 & $\pm\!$83 & 5.089
      & $\pm\!$0.216 & $57.2$ & $\pm\!$31.0 & Blue straggler \\
      candBHB21 & 2459177.77718009 & 8350 & $\pm\!$159 & 3.063
      & $\pm\!$0.091 & $67.8$ & $\pm\!$9.2 & BHB \\
      candBHB22 & 2459199.8012333 & 7513 & $\pm\!$59 & 2.855
      & $\pm\!$0.270 & $53.7$ & $\pm\!$11.5 & BHB \\
      candBHB23 & 2459176.79133185 & 8752 & $\pm\!$354 & 3.027
      & $\pm\!$0.161 & $55.1$ & $\pm\!$12.4 & BHB \\
      \multirow{2}{*}{candBHB24} & 2459196.80650984 & 8583 & $\pm\!$262
      & 3.380 & $\pm\!$0.105 & $72.9$ & $\pm\!$15.3 & BHB \\
      & 2459195.81788969 & 8935 & $\pm\!$358 & 3.329
      & $\pm\!$0.211 & $69.9$ & $\pm\!$14.2 & BHB \\
      & 2459197.8023447 & 10415 & $\pm\!$852 & 3.873 & $\pm\!$0.367
      & $75.7$ & $\pm\!$15.2 & BHB \\
      \multirow{2}{*}{candBHB25} & 2459198.81629641 & 11118 & $\pm\!$2
      & 6.465 & $\pm\!$0.002 & $124.3$ & $\pm\!$29.1 & WD \\
      & 2459225.71684968 & 10787 & $\pm\!$nan & 6.456 & $\pm\!$nan
      & $99.6$ & $\pm\!$28.5 & WD \\
      \hline
    \end{tabular}
    \tablefoot{A small subset of the returned parameters was chosen
               because of the expected difficulty in obtaining reliable
               estimates of [Fe/H], [$\alpha$/Fe], and $v\sin\ i$ from
               low-resolution optical spectra of hot stars with strong
               Balmer lines that blend some of the metallicity-%
               sensitive lines
               \citep[%
                 see also noted limitations of
                 \textsc{\lowercase{RVSpecFit}} in
               ][]{koposov_et_al_arxiv2407_06280}.
               `MAP' is used here to refer to the maximum a posteriori
               estimations of the quantities within
               \textsc{\lowercase{RVSpecFit}}.}
    \label{tab:results_cand_bhb_stars_eso_p106prgs}
  \end{table*}

  The findings indicate that the purity of the sample in question
  (80\,per\,cent) corresponds to the anticipated range for the
  photometric selection of candidate BHB stars, as outlined in
  \citet{starkenburg_et_al2019}. This is evident in the middle panel of
  figure 5 in \citet{starkenburg_et_al2019}, which assumes a one-to-one
  ratio of BHB to blue straggler stars in the outer halo. 

  \subsection{Blue Horizontal Branch stars in Sagittarius spur region}
  
\begin{figure}
    \includegraphics{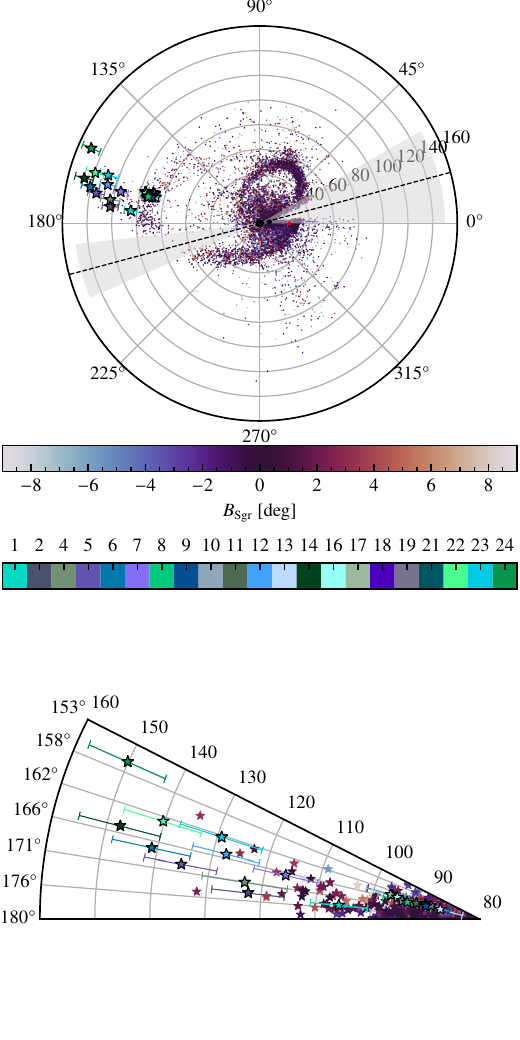}
    \caption{Sagittarius stream according to a sample of candidate
             members in sample of (candidate) variable RR Lyrae stars
             by \citet{hernitschek_et_al2017} in one projection across 
             longitude of the Sagittarius stream coordinate system as
             defined by \citet{vasiliev_et_al2021}
             \citep[implemented in \textsc{\lowercase{Gala}}][]{gala},
             and heliocentric distance. For each (candidate) RR Lyrae 
             we colour-code the marker by the latitude in the same 
             Sagittarius stream coordinate system. The top half of the 
             mapped Sagittarius stream at between 135\degr and 180\degr
             longitude (see lower panel for enlarged plot of this part),
             includes the BHB stars in our sample plotted with distinct
             colours based on their id (in the label of the color scheme
             we only show the number at the end of id).}
    \label{fig:adapt_sesar_et_al2017b_figure1low_panel}
  \end{figure}

  In the following, we will present the kinematics of the stars 
  classified as BHB stars. 
  Figure \ref{fig:adapt_sesar_et_al2017b_figure1low_panel} presents a 
  map of the Sagittarius stream according to a sample of candidate 
  members in sample of variable RR Lyrae stars by 
  \citet{hernitschek_et_al2017} in one projection across longitude of 
  Sagittarius stream coordinate system as defined by 
  \citet{vasiliev_et_al2021} and heliocentric distance. We also show 
  the BHB stars in the Sagittarius spur region in 
  \autoref{fig:adapt_sesar_et_al2017b_figure1low_panel}. To estimate 
  the heliocentric distances of the BHB stars, numerous studies of the 
  Galactic stellar halo traced by BHB stars have utilised the 
  fourth-order polynomial absolute magnitude relation derived from the 
  SDSS DR6 photometry of BHB stars in ten star clusters, as presented 
  by \citet{an_et_al2008}. Unlike \citet{deason_et_al2011},
  \citet{barbosa_et_al2022} argue that 
  \begin{equation}
    M_{g_{\text{SDSS},0},\text{BHB}}%
    \left((g\!-\!r)_{\text{SDSS},0}\right)%
    \!=\! \frac{0.178}{0.537\!+\!(g\!-\!r)_{\text{SDSS},0}}
    \label{equ:abs_sloan_g_mag_bhb_rel}
  \end{equation}
  results in smaller uncertainties for the absolute magnitude of BHB
  stars in Sloan $g_0$ when revisiting the approximation by
  \citet{deason_et_al2011} with more data of BHB stars in star clusters
  and better constraints on the distances of the star clusters. In both
  \citet{deason_et_al2011} and \citet{barbosa_et_al2022}, the intrinsic
  spread in the absolute magnitude/colour relation is of 0.1\,mag. This
  spread takes into account the metallicity-dependence of the 
  colour-absolute magnitude relation in case of 
  \citet{deason_et_al2011}. In contrast, \citet{barbosa_et_al2022} find 
  that \autoref{equ:abs_sloan_g_mag_bhb_rel} does not have any 
  significant metallicity-dependence. 

  We adopt \autoref{equ:abs_sloan_g_mag_bhb_rel} to calculate the 
  heliocentric distances of the BHB stars while noting that the output 
  obtained from the \citet{deason_et_al2011} polynomial are consistent 
  with the distances resulting from 
  \autoref{equ:abs_sloan_g_mag_bhb_rel} within their uncertainties. It 
  is important to stress here that not all the stars used here have an 
  SDSS photometric measurement, in particular for the stars selected 
  from the \citet{thomas_et_al2018} catalogue. In that case, the colour 
  in the SDSS photometric system is computed from the PS1 measurement 
  using the equation~6 of \citet{thomas_et_al2018}. To enable the 
  computation of
  $\text{SDSS\ }g_0 \!-\!  M_{\text{SDSS\ }g_0,\text{BHB}}$,
  $g _ {\text{SDSS},0}$ were approximated with $g _ {\text{PS},0}$. The 
  uncertainties on the distances were computed by adding in quadrature 
  the uncertainties due to the photometric uncertainties in the 
  \textit{g} and \textit{r} bands to the intrinsic scatter of the 
  colour-absolute magnitude relation. This results in uncertainties of
  typically 5\,per\,cent.

  \section{Results}\label{sec:results} 
  
  \subsection{Comparison between observations and N-body simulations
              of stars in the distant part of the Sagittarius tidal
              stream}%
  \label{subsec:dist_trail_arm_sgr_stream}

  We explore the relationship between heliocentric distance and
  (corrected) LOS velocity for the BHB stars in our sample relative to 
  predictions of N-body simulations of the formation and evolution of
  the Sagittarius stream.

  \subsubsection{Radial velocity structure of Blue Horizontal Branch
                 stars}%
  
\begin{figure*}
    \centering
  	\includegraphics{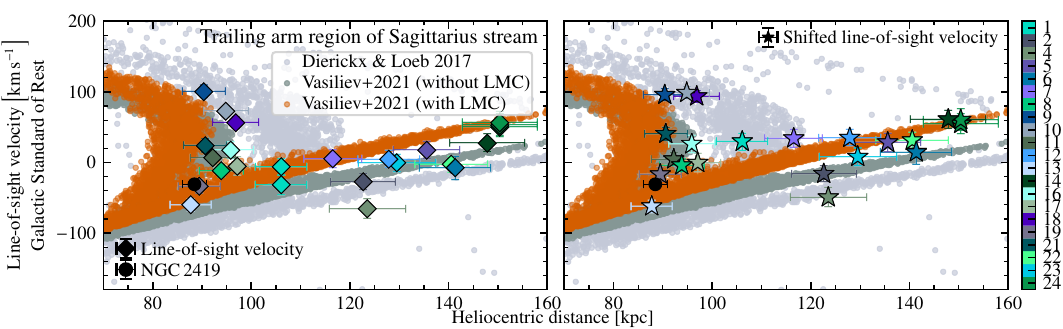}
  	\caption{Comparison of the observational data on the LOS velocities
             in the Galactic Standard of Rest of the BHB stars (diamonds
             and stars) and predictions from the
             \citet{dierickx_loeb2017} (lightgreyish points) and
             \citet{vasiliev_et_al2021} without (darkgreyish points)
             and with LMC (orange points) models in the same angular
             range across the stream. Left panel: big diamond markers
             indicate LOS velocities in Galactic Standard of Rest of BHB
             stars estimated with \textsc{\lowercase{RVSpecFit}} from
             the FORS2 spectra. Right panel: stars represent the same
             velocities with an applied shift as reported in Figs.
             \ref{fig:comparison_los_vel_estimates_vel_sts_main} 
             and
             \ref{%
               fig:comparison_los_vel_estimates_vel_sts_appendix%
             }. 
             The plot shows also the position of globular cluster 
             NGC\,2419 (filled circle) in this projection of phase 
             space. The LOS velocity of NGC\,2419 comes from
             \citet{vasiliev_baumgardt2021} assuming also the right
             ascension and declination provided in the
             \citet{vasiliev_baumgardt2021} catalogue while we use the 
             heliocentric distance of NGC\,2419 from 
             \citet{baumgardt_vasiliev2021}.}
  	\label{fig:dist_sgr_stream_vel_struct_main}
  \end{figure*}

  Figure \ref{fig:dist_sgr_stream_vel_struct_main} examines the radial 
  velocity structure (stars and diamonds for results that are velocity-%
  corrected and not, respectively) of the apocenter of the trailing arm
  and spur feature of the Sagittarius stream as traced by our sample of
  BHB stars\footnote{For the purpose of Galactic Standard of Rest LOS
            velocity calculation, we adopted the definition of the 
            Galactocentric coordinate system as also used in 
            \citet{vasiliev_et_al2021} that is the standard in 
            \textsc{\lowercase{Astropy}}
            \citep{astropy2013, astropy2018, astropy2022}.
            This definition uses the J2000 International Celestial 
            Reference System coordinates
            ($\mathrm{R.A.} \!=\! 17^\mathrm{h}45^\mathrm{m}37\fs224$,
            $\mathrm{Decl.} \!=\! -28\degr56\arcmin10\farcs23$) of the 
            Galactic centre noted in \citet{reid_brunthaler2004}, a
            $\mathrm{distance} \!=\! 8.122$\,kpc between the Sun and 
            Galactic centre as constrained by
            \citet{gravity_collab2018},
            $\mathrm{height} \!=\! 20.8$\,pc of the Sun above the 
            Galactic mid-plane, and
            $(12.9, 245.6, 7.78)\,\mathrm{km}\,\mathrm{s}^{-1}$
            \citep{drimmel_poggio2018} as the 3D Solar velocity 
            relative to the Galactic centre.}.
  The constraints on the kinematics of the distant part of the trailing 
  arm of the Sagittarius stream that is traced by our sample of BHB 
  stars are compared to the present-day data of stellar particles in 
  three simulations of the infall of the Sagittarius dwarf galaxy.
  
  The three simulations we consider are by
  \citet[][predictions in 
           \autoref{fig:dist_sgr_stream_vel_struct_main} are shown with 
           grey points]%
  {dierickx_loeb2017}% 
  \footnote{The data was gathered from 
            \url{mdierick.github.io/Sgr\_stars\_kin\_data.csv}. The
            LOS velocities of the particles in their simulation are 
            already in the Galactic Standard of Rest with a slightly 
            different definition of the Galactocentric coordinate 
            system.}
  and
  \citet[][one takes into account the effects of the LMC and the other 
           does not]%
  {vasiliev_et_al2021}.
  Data of the stellar particles in the N-body simulations by 
  \citet{vasiliev_et_al2021} are represented by orange and dark-greyish 
  points in 
  \autoref{fig:dist_sgr_stream_vel_struct_main} for the case of the
  simulation without and with the LMC, respectively%
  \footnote{We use the same procedure for the computation of the LOS 
            velocities in the Galactic Standard of Rest for the 
            stellar particles in the \citet{vasiliev_et_al2021} model as
            for the BHB stars.}.
  The main differences between these simulations are as follows:
  \begin{itemize}
    \item no LMC perturbations in \citet{dierickx_loeb2017} simulation.
    \item \citet{hernquist1990} model for gravitational potential of
    Galactic bulge with total mass of
    $1.25\!\times\!10^{10} \, \text{M}_\odot$ in
    \citet{dierickx_loeb2017} versus a exponentially-truncated,
    spheroidal power law model with total mass of 
    $1.2\!\times\!10^{10} \, \text{M}_\odot$ in
    \citet{vasiliev_et_al2021} for the bulge density.
    \item exponentially declining gravitational potential of disk with
    total mass of $8.125\!\times\!10^{10} \, \text{M}_\odot$ in
    \citet{dierickx_loeb2017} versus isothermal sheet with exponential
    radial variation for the disk density and total mass of
    $5\!\times\!10^{10} \, \text{M}_\odot$ in
    \citet{vasiliev_et_al2021}.
    \item spherical \citet{hernquist1990} model for gravitational
    potential of Galactic halo with total mass of
    $1.25\!\times\!10^{12} \, \text{M}_\odot$ in
    \citet{dierickx_loeb2017} versus a exponentially-truncated,
    \citet{h_zhao1996} model with varying flattening and orientation in
    \citet{vasiliev_et_al2021} for halo density.
    \item \citet{dierickx_loeb2017} adopt the same models assumed to
    describe the gravitational potential of the Milky Way for the
    Sagittarius progenitor where bulge, disk, and halo have total masses
    of $5.2\!\times\!10^{8} \, \text{M}_\odot$,
    $7.8\!\times\!10^{8} \, \text{M}_\odot$, and
    $1.3\!\times\!10^{10} \, \text{M}_\odot$ while
    \citet{vasiliev_et_al2021} assume the King model for the stellar
    density of the Sagittarius progenitor and a more extended,
    spherical, cored dark matter (DM) halo of mass of
    $3.6\!\times\!10^9\,\text{M}_\odot$ that is gradually tidally
    disrupted with a mass loss described by a piecewise linear 
    function. 
  \end{itemize}
  
  In \autoref{fig:dist_sgr_stream_vel_struct_main} there is a clear
  trend of decreasing Galactic Standard of Rest LOS velocities
  beyond heliocentric distance $D_\odot \!\simeq\! 120$\,kpc for the
  data of the final (present-day) snapshot of the
  \citet{dierickx_loeb2017} and \citet{vasiliev_et_al2021} N-body
  simulations of the formation and evolution of the Sagittarius stream.
  To a lesser extent we see a similar trend in the BHB stars (both the
  estimates of the LOS velocity without and with shifts to correct the
  systematics outlined in \autoref{subsec:accuracy_los_vels} in the left
  and right panel, respectively). The correlation between heliocentric
  distance and Galactic Standard of Rest LOS velocity for
  $D_\odot \!\gtrsim\! 120$\,kpc is interesting because these are the
  stars expected in the spur feature of the trailing arm of the
  Sagittarius stream and the fact that the BHB stars trace such a 
  feature qualitatively confirms its existence. The models shown in 
  \autoref{fig:dist_sgr_stream_vel_struct_main} and also
  \citet{fardal_et_al2019}\footnote{The data for this model is not 
                                    publicly available.}
  have significant different predictions for the velocity structure of 
  this feature.
  
  In both panels of \autoref{fig:dist_sgr_stream_vel_struct_main}, the
  trend for the spur feature as traced by the BHB stars follows the 
  models from \citet{vasiliev_et_al2021} more closely. Depending on
  whether the tentative correction is applied (right panel), it follows
  more the \citet{vasiliev_et_al2021} model with the LMC. However, we
  also note that in both cases (without and with the correction) the
  apocentre data points fit better with the \citet{vasiliev_et_al2021}
  model without the LMC. 
  
  In \autoref{fig:dist_sgr_stream_vel_struct_main}, we note that the
  shifts applied to the LOS velocities of the BHB stars based on the
  values in \autoref{fig:comparison_los_vel_estimates_vel_sts_main}
  (from the left to the right panel) are not significant compared to the
  large velocity range that we study. Second order variations in the LOS
  velocity measurements such as due to the flexure therefore
  reassuringly do not affect our conclusions. Over half of the BHB 
  stars surveyed in our sample could be part of the apocentre of the 
  trailing arm which is visible in
  \autoref{fig:dist_sgr_stream_vel_struct_main} below 120\,kpc as the
  arc charactertistic of the apocentres where stars turn around and have
  smaller and smaller absolute LOS velocities. Interestingly,
  the position of the apocentre of the trailing arm according to our
  data was observed to be systematically off of any the models presented
  in \autoref{fig:dist_sgr_stream_vel_struct_main}. In summary, the 
  constrained LOS velocity structure of the apocentre of the trailing 
  arm and the spur beyond it of the Sagittarius stream with our sample 
  of BHB stars is informative to constrain current models of the 
  Sagittarius stream and also in turn the Galactic gravitational 
  potential at these distances.
  
  We note that \citet{jing_li_et_al2023} present a set
  of (candidate) M-type giant stars that also reaches the spur region
  of the distant Sagittarius stream trailing arm but with very limited
  accuracy in their distances (the uncertainties of the photometric,
  heliocentric distances $\Delta D_\odot \!\gtrsim\! 30$\,kpc at
  $D_\odot \!\gtrsim\! 70$\,kpc, hence 3-7 times larger than we find
  for the BHB stars, see also 
  \autoref{appendix_sec:cand_m_type_giants}). Our sample traces this
  region combining accurate distances and velocity space for the first 
  time. In the next two sections we discuss more quantitatively how we
  can constrain models of the formation and evolution of the 
  Sagittarius stream and the shape of the Galactic gravitational 
  potential at the radii that our stars cover.
  
  \subsubsection{Distance rescaling of Sagittarius apocentre and
                 spur in \citet{vasiliev_et_al2021} simulations}%
  \label{sec:distance_rescaling}
  
  As discussed above, we find there is a difference in the mapped
  velocity structure of the BHB stars in the most distant part of the 
  trailing arm of the Sagittarius stream and predictions from the 
  \citet{dierickx_loeb2017} and \citet{vasiliev_et_al2021} models 
  (see \autoref{fig:dist_sgr_stream_vel_struct_main}). Of the two 
  models, the latter resembles the trend expressed by our 
  data in this study more. This model is also a bit closer to the
  observed position of the apocentre of the trailing arm of the
  Sagittarius stream in the 2D projection of heliocentric distance and
  LOS velocity. Nonetheless, this apocentre turn-around seems somewhat
  offset in distance in the model predictions compared to our data
  points. A possible explanation for this might be that the enclosed 
  Galactic mass at the apocentre of the trailing arm (here 100\,kpc) 
  found by \citet{vasiliev_et_al2021} of
  $(5.6\!\pm\!0.4)\!\times\!10^{11} \, \mathrm{M}_\odot$ is lower. In 
  consideration of the fact that the \citet{vasiliev_et_al2021} model 
  accurately reproduces the majority of the Sagittarius stream at 
  closer Galactic distances, it is also proposed that the assumed 
  outer slope $\beta$ of the halo density profile by 
  \citet[
    $\rho \!\propto\!(R/\text{rscale})^{-\gamma}%
    \left(1\!+\!(R/\text{rscale})^\alpha\right)%
    ^{(\gamma-\!-\!\beta)/\alpha}%
    \exp\left(-(R/200\,\text{kpc})^2\right)$
    and
    $R \!\equiv\!(pq)^{1/3}\sqrt{X^2\!+\!(Y/p)^2\!+\!(Z/q)^2}$
  ]{vasiliev_et_al2021}
  is not an accurate description of the true potential at these 
  (relatively unconstrained) distances.
  
\begin{figure}
    \centering
    \includegraphics{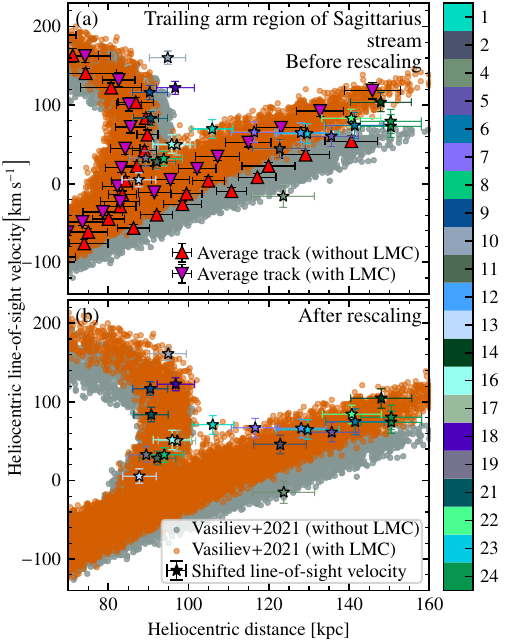}
    \caption{Rescaling of the average track of stellar particles in
             heliocentric distance and LOS velocity in final (present-%
             day) snapshots of N-body simulations of Sagittarius stream
             by \citet{vasiliev_et_al2021} without and with
             perturbations by the LMC across heliocentric distance. (a)
             Copy of right panel of
             \autoref{fig:dist_sgr_stream_vel_struct_main} except 
             replacing Galactic Standard of Rest by heliocentric LOS
             velocities, plotting data of BHB stars as shown in the
             right panel of
             \autoref{fig:dist_sgr_stream_vel_struct_main}, removing
             data of the particles in the \citet{dierickx_loeb2017}
             model, and adding average tracks of the data of the stellar
             particles in the \citet{vasiliev_et_al2021} N-body
             simulations. (b) heliocentric 2D phase space data of
             stellar particles of N-body simulations presented in
             \citet{vasiliev_et_al2021} after rescaling to match it to
             the shown constraints from the BHB stars.}
    \label{fig:dist_sgr_stream_vel_struct_rescaling}
  \end{figure}

  To investigate how $\beta$ and in turn the enclosed mass of the MW
  within 100\,kpc has to be adapted to match our constraints, we rescale
  across heliocentric distance the data of the stellar particles
  associated to the trailing arm region of the 
  Sagittarius stream in the presented N-body simulations by 
  \citet{vasiliev_et_al2021} without and with the LMC (cf. with 
  \autoref{fig:dist_sgr_stream_vel_struct_main}). Average tracks were 
  generated for all data of the \citet{vasiliev_et_al2021} models in 
  \autoref{fig:dist_sgr_stream_vel_struct_main} (stellar particles at
  $-230\,\degr \!<\! \Lambda _ \text{Sgr} \!<\! -160\,\degr$ in final,
  present-day snapshots of the N-body simulations). These tracks are 
  shown in \autoref{fig:dist_sgr_stream_vel_struct_rescaling}, where the
  top panel points were obtained by grouping the 500 closest data 
  points together in heliocentric distance and velocity space using 
  K-means clustering as implemented in 
  \textsc{\lowercase{Scikit-learn}} \citep{scikit-learn} and finding 
  the standard deviations that describe each group/cluster of data 
  points. 

  Chi-square tests were used to compare the average tracks to four
  considered cases of some representation of the data in
  Figs. \ref{fig:dist_sgr_stream_vel_struct_main} and
  \ref{fig:dist_sgr_stream_vel_struct_rescaling} excluding outlier
  candBHB4%
  \footnote{No significant difference were found between results from 
            these tests with and without the data of candBHB4. Its 
            classification is uncertain (see 
            \autoref{tab:results_cand_bhb_stars_eso_p106prgs}).}.
  These four cases are: 
  \begin{itemize}
    \item (B22, Heliocentric) with heliocentric distance of BHB stars 
    estimated from \autoref{equ:abs_sloan_g_mag_bhb_rel} and LOS
    velocities of BHB stars (data represented in the left panel of
    \autoref{fig:dist_sgr_stream_vel_struct_main}).
    \item (B22, Corrected) with heliocentric distance of BHB stars 
    estimated from \autoref{equ:abs_sloan_g_mag_bhb_rel} and corrected
    LOS velocities of BHB stars (data represented in the right panel of
    \autoref{fig:dist_sgr_stream_vel_struct_main}).
    \item (D11, Heliocentric) heliocentric distance of BHB stars
    estimated from \citet{deason_et_al2011} relation and LOS velocities
    of BHB stars.
    \item (D11, Corrected) with heliocentric distance of BHB stars 
    estimated from \citet{deason_et_al2011} relation and corrected LOS 
    velocities of BHB stars.
  \end{itemize}
  Subsequently, we sample the natural logarithm of the posterior of the 
  rescaling factor $f$ given the data that we calculated from the 
  natural logarithm of a Gaussian likelihood (equivalent to 
  approximately $\chi^2$) of the difference between the data of the BHB 
  stars and the average tracks to find the optimal value of $f$. The 
  uncertainties of the average tracks and observational data were added 
  in quadrature, and we adopt a uniform prior for $f$ over 0.1 and 10.0.
  This sampling across Markov chains with a Monte Carlo algorithm and
  optimisation was achieved with \textsc{\lowercase{emcee}}
  \citep{foreman-mackey_et_al2013} using ten walkers and 1,000 steps.
  \autoref{tab:rescaling_factor} lists the results of the rescaling
  (median and 16 and 84 quantiles of sampled posterior distribution of
  $f$). We note that none of the rescaling experiments within the group
  of cases including the LMC influence (`E', `F', `G', and `H') as well
  as within the group of cases without the LMC influence (`A', `B',
  `C', and `D') resulted in differences that were statistically
  significant. This illustrates again that - considering the main
  results presented in this paper - the possible flexure corrections of
  the radial velocities are not that influential and, moreover, that the
  choice of distance calibration for the BHB stars does not alter the
  results significantly. However, as maybe expected, the difference for
  $f$ between the two groups (with and without the LMC perturbation) is 
  statistically significant. For the remainder of this work, case `D' 
  (B22, Corrected, No) and `H' (B22, Corrected, Yes) are labeled as our 
  fiducial cases with and without the LMC. 
  
  Panel (b) of 
  \autoref{fig:dist_sgr_stream_vel_struct_rescaling} compares the 
  rescaling of the 2D heliocentric phase space (distance, LOS velocity) 
  data of the stellar particles at
  $-230\,\degr \!<\! \Lambda _ \text{Sgr} \!<\! -160\,\degr$ along
  the Sagittarius stream in the present-day snapshots of the N-body
  simulations without ($f \!=\! 1.05^{+0.02}_{-0.02}$ for fiducial case)
  and with the LMC ($f \!=\! 1.11^{+0.02}_{-0.02}$ for fiducial case) as
  performed by \citet{vasiliev_et_al2021} for the fiducial cases. The
  results indicate that the rescaled \citet{vasiliev_et_al2021} model 
  with the LMC fits the data very well. Indeed, this supports a number
  of previous works that demonstrate the effects of the LMC on the MW 
  and the overdensities in it
  \citep[e.\,g.,][]{%
    erkal_et_al2019, garavito-camargo_et_al2019, shipp_et_al2019, 
    petersen_penarrubia2021, vasiliev_et_al2021, 
    garavito-camargo_et_al2021, lilleengen_et_al2023, koposov_et_al2023,
    v_chandra_et_al_arxiv2406_01676, bystroem_et_al_arxiv2410_09149}
  
  \begin{table}
    \centering
    \caption{Differences between the data of the velocity structure of 
             the BHB stars at the apocenter and spur feature of the
             trailing arm of the Sagittarius stream to which the data
             of the final (present-day) snapshots of the N-body
             simulations of the formation and evolution of the
             Sagittarius stream in the pre- or absence of the LMC by
             \citet{vasiliev_et_al2021} were scaled along heliocentric
             distance to match it.}
    \begin{tabular}{lllll}
      \hline
      Label & Distance & LOS velocity & LMC & $f$ \\
      (1) & (2) & (3) & (4) & (5) \\
      \hline
      A & D11 & Heliocentric & No & 1.04$^{+0.02} _ {-0.04}$ \\
      B & B22 & Heliocentric & No & 1.05$^{+0.02} _ {-0.03}$ \\
      C & D11 & Corrected & No & 1.03$^{+0.02} _ {-0.02}$ \\
      D & B22 & Corrected & No & 1.05$^{+0.02} _ {-0.02}$ \\
      E & D11 & Heliocentric & Yes & 1.12$^{+0.02} _ {-0.02}$ \\
      F & B22 & Heliocentric & Yes & 1.15$^{+0.02} _ {-0.02}$ \\
      G & D11 & Corrected & Yes & 1.10$^{+0.02} _ {-0.02}$ \\
      H & B22 & Corrected & Yes & 1.11$^{+0.02} _ {-0.02}$ \\
      \hline
    \end{tabular}
    \tablefoot{The second column is the abbreviation for the equation
               that we use to calculate the absolute magnitude in Sloan
               $g _ \text{SDSS}$ of the sample BHB stars and as a result
               distance modulus and heliocentric distance of the stars.
               Wheras `D11' refers to \citet{deason_et_al2011}, `B22'
               refers to \citet{barbosa_et_al2022}. (3) abbreviates
               which LOS velocity of each BHB star in the set that we
               study we consider for the rescaling. Firstly, the term
               `Heliocentric' refers to the heliocentric LOS velocity
               without a shift to correct for systematics as estimated
               from the velocity standard stars (see 
               \autoref{fig:comparison_los_vel_estimates_vel_sts_main} 
               or 
               \autoref{%
                 fig:comparison_los_vel_estimates_vel_sts_appendix}
               and left panel of 
               \autoref{fig:dist_sgr_stream_vel_struct_main}).
               Secondly, `Corrected' refers to the corrected
               (heliocentric) LOS velocities (see right panel of 
               \autoref{fig:dist_sgr_stream_vel_struct_main}). (4) 
               indicates whether this is the data from the N-body 
               simulation by \citet{vasiliev_et_al2021} with (`Yes') or 
               without (`No') perturbations by the LMC.}
    \label{tab:rescaling_factor}
  \end{table} 
  
  \subsection{Milky Way mass profile}\label{subsec:mwprofile}
  
  Given the similarities between the rescaled \citet{vasiliev_et_al2021}
  model with the LMC and the observed Sagittarius stream, it is possible
  to use this model to re-evaluate the MW's mass within the Sagittarius 
  apocentre radius ($\simeq 100$~kpc). In this section, we present a 
  first analysis in this direction to provide an estimate of the 
  (magnitude of) change in mass implied, but we note that a full and 
  detailed re-modeling is beyond the scope of this work.   
  
  The rescaling factor derived in \autoref{sec:distance_rescaling} is
  computed such that the heliocentric LOS velocities in the model match
  the observed, heliocentric velocities at the rescaled Galactocentric
  positions\footnote{
    While the rescaling results presented in
    \autoref{sec:distance_rescaling} are across heliocentric distances,
    we use here the effect of the rescaling of the heliocentric
    distances on the Galactocentric positions and furthermore note that
    rescaling in Galactocentric distances instead yields consistent
    results compared to the entries in \autoref{tab:rescaling_factor}}.
  In this analysis, we exclusively utilise the LOS velocity. Note that
  proper motions are not considered because they are not available for
  all stars. 

  In our approach, we assume that the Sagittarius stream model from
  \citet{vasiliev_et_al2021} provides a good representation of the total
  energy distribution at every position along the stream. However, as
  shown in \autoref{fig:dist_sgr_stream_vel_struct_rescaling}, we have
  seen that the model needs to be stretched to match the distance and
  velocity distribution of the data. Therefore, we stretch the
  Galactocentric coordinates of the Sagittarius stream model of
  \citet{vasiliev_et_al2021} (with the LMC) according to the rescaling
  factor found in \autoref{sec:distance_rescaling}, but leave the 3D
  velocities of the particles unchanged. Adjusting the particles in the
  \citet{vasiliev_et_al2021} simulation according to this rescaling
  factor provides direct access to the gravitational potential
  ($\Phi$) of each star given that 
  $\mathrm{E} \!\equiv\! \Phi + 0.5\ \vec{V}^2$, where $\mathrm{E}$ is
  the total energy and \vec{V} the Galactocentric velocity of the star.

  Consequently, we adjust the MW DM halo profile such that the
  gravitational potential with the new profile,
  $\Phi_\text{MW,new}(\vec{R})$, of a star at the rescaled
  Galactocentric position $\vec{R'}$ matches the potential of the
  \citet{vasiliev_et_al2021} model, $\Phi_\text{MW,Vasiliev+2021}$, at
  Galactocentric position $\vec{R}$ before rescaling.
  
  Moreover, given that the original (non-rescaled)
  \citet{vasiliev_et_al2021} model accurately reproduces most of the
  Sagittarius stream at shorter Galactic distances, we only adjust the
  outer slope of the MW DM halo. Specifically, we assume that the
  triaxiality, twisting, and baryonic distribution remain unchanged
  compared to the MW model proposed by \citet{vasiliev_et_al2021}%
  \footnote{Our adjustment of the MW DM profile incorporates
            uncertainties in these parameters.}.
  In particular, we note that we thereby assume that the gravitational
  perturbation of the LMC remains unchanged at the rescaled position.
  This assumption is not entirely correct, as for a same fixed mass of
  the LMC as found by \citet{vasiliev_et_al2021}, its perturbation on
  the apocentre of the Sagittarius stream in the rescaled model should
  be less important than at the original positions of the
  \citet{vasiliev_et_al2021} model, given that the rescaled position are
  further away from the LMC than the original position. At the same
  time, the mass ratio of the LMC and MW will be more comparable with a
  lower mass MW. Properly accounting for these perturbations would
  require rerunning the simulation of \citet{vasiliev_et_al2021}, as the
  LMC directly exerts a gravitational force on the Sagittarius stream
  but also modifies the shape of the DM halo
  \citep{garavito-camargo_et_al2019, petersen_penarrubia2021}. However,
  we stress this will be a secondary effect due to the relatively small
  changes in distance compared to the original model. 

  We impose the boundary condition on the gravitional potential that 
  the gravitational potential at the Galactocentric location of the 
  Sagittarius dwarf spheroidal galaxy remnant,
  $\vec{R}_\text{Sgr} \!=\! (17.9, 2.6, -6.6)$\,kpc, should be the same 
  between the original \citet{vasiliev_et_al2021} model and the 
  adjusted DM profile:
  \[%
  \Phi_\text{MW,new}\left(\vec{R}_\text{Sgr}\right)%
  \!=\! \Phi_\text{MW,Vasiliev+2021}\left(\vec{R}_\text{Sgr}\right).%
  \]

  Thus, revising the enclosed Galactic mass within 100\,kpc and by
  extension exploring the changes in the outer slope $\beta$ of the MW
  DM halo involves minimising the following least-squares function:
  \begin{equation}
    \chi^2 \!=\! \sum_{i=1}^N%
    \left[\Phi_\text{MW,new}\left(\vec{R'}_i\right)%
    \!-\! \Phi_\text{MW,Vasiliev+2021}\left(\vec{R}_i\right)%
    \!-\! \text{norm}\right]^2,
    \label{equ:likelihood_func}
  \end{equation}
  where $\vec{R'}_i$ represents the rescaled, Galactocentric positions
  along the stream and $\text{norm}$ accounts for normalisation
  adjustments. The subscript $i$ represents the $N$ particles in the
  distant trailing arm region of the Sagittarius stream 
  $\left(%
    -230\,\degr \!<\! \Lambda _ \text{Sgr} \!<\! -160\,\degr\right)$
  from the \citet{vasiliev_et_al2021} simulation. In the region of the 
  Sagittarius stream with the data of the BHB stars, the positional,
  Galactocentric vector ($X _ i$, $Y _ i$, $Z _ i$) is represented by
  $\vec{R}_i$ for the ith particle. $\Phi_\text{MW,new}$ is calculated
  using the stretched, Galactocentric $X' _ i$, $Y' _ i$, and $Z' _ i$
  positions found from the rescaling described in
  \autoref{subsec:dist_trail_arm_sgr_stream} to yield the potential in
  $\vec{R}' _ i \!=\! \left(X'_i, Y'_i, Z'_i\right)$ for the ith 
  particle. All potentials are computed with the Action-based Galaxy 
  Modelling Architecture code
  \citep[\textsc{\lowercase{AGAMA}},][]{vasiliev2018, vasiliev2019a}
  version 1.0.
  
  We employ a Monte Carlo technique to sample from the unknown 
  distribution of the enclosed MW mass within 100 kpc and $\beta$ by
  repeating the minimisation of \autoref{equ:likelihood_func} to find
  the optimal value of the mass and $\beta$ with 1,000 realisations of
  $f$ and thus $\vec{R}' _ i$ and the other parameters of the halo
  density profile model by \citet{vasiliev_et_al2021}. In the Monte
  Carlo method, for each realisation we vary the rescaling factor
  $f \!=\! 1.11 \!\pm\! 0.02$, scale radius rscale, inner slope
  $\gamma$, and transition steepness $\alpha$ of the halo density
  profile model by \citet{vasiliev_et_al2021} (Eugene Vasiliev, private
  communication) and the other parameters of the model by
  \citet{vasiliev_et_al2021} that quantify the triaxiality and twisting
  of the halo \citep[see figure 14 in][]{vasiliev_et_al2021} within
  their uncertainties. We use split normal distributions to approximate
  the distributions of the parameters of the halo density profile model
  by \citet{vasiliev_et_al2021}. We do not take into account here that 
  some of the parameters of the \citet{vasiliev_et_al2021} model are 
  correlated with $\beta$, but treat them as independent. The 
  methodology employed here is relatively straightforward and, it must 
  be acknowledged, somewhat simplistic. It is assumed that the energy 
  distribution for the Sagittarius progenitor provided by the 
  \citet{vasiliev_et_al2021} model is reliable, given that it already 
  closely matches the majority of the stream. However, it is our 
  contention that it can provide a satisfactory initial approximation 
  of the local potential before undertaking further, more comprehensive 
  simulations in future work. 
    
  The results of the sampling show that
  $\beta \!=\! 2.47^{+0.26}_{-0.21}$ can lead to a position of the
  apocentre of the trailing arm of the Sagittarius stream in the
  \citet{vasiliev_et_al2021} that is in line with our constraints based
  on the BHB stars. This value of $\beta$ is slightly higher, but within
  the uncertainties of the $\beta$ value reported by
  \citet[$\beta \!=\! 2.43^{+0.16}_{-0.13}$; Eugene Vasiliev, private
         communication]%
  {vasiliev_et_al2021}.
  Consequently, the enclosed mass at the apocentre of the trailing arm
  (here $\sim$100\,kpc) by \citet{vasiliev_et_al2021} is lowered to
  $(5.3\!\pm\!0.4)\!\times\!10^{11}\,\mathrm{M}_\odot$, again consistent
  within the uncertainties of \citet{vasiliev_et_al2021} who report
  $M(<100\,\text{kpc})%
  \!=\! (5.6\!\pm\!0.4)\!\times\!10^{11}\,\mathrm{M}_\odot$.

\begin{figure}
    \centering
    \includegraphics{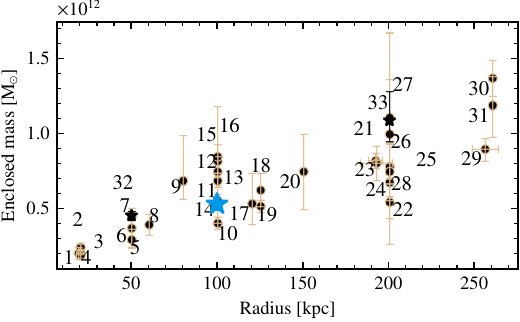}
    \caption{Adapted from \citet{bobylev_baykova2023} showing estimates 
             of enclosed mass of Milky Way at different radii together
             with our constraints with the (blue) star symbol at 100\,%
             kpc. As in the diagram in \citet{bobylev_baykova2023} that 
             we adapted here, with the exceptions of 32 and 33 that are
             new estimates by \citet{ibata_et_al2024} and are 
             highlighted here, the annotated numbers for each data 
             point correspond to  1--\citet{kuepper_et_al2015}, 2--%
             \citet{malhan_ibata2019}, 3--\citet{prudil_et_al2022}, 4--%
             \citet{posti_helmi2019}, 5--\citet{williams_et_al2017},
             6--\citet{ablimit_zhao2017}, 7--\citet{williams_evans2015},
             8--\citet{xue_et_al2008}, 9--\citet{gnedin_et_al2010},
             10--\citet{gibbons_et_al2014}, 11--%
             \citet{vasiliev_et_al2021}, 12--%
             \citet{jeff_shen_et_al2022}, 13--%
             \citet{correa_magnus_and_vasiliev2022}, 14--%
             \citet{eadie_juric2019}, 15--\citet{mcmillan2017}, 16--%
             \citet{vasiliev2019b}, 17--\citet{battaglia_et_al2005}, 
             18--\citet{eadie_et_al2017}, 19--\citet{eadie_harris2016},
             20--\citet{deason_et_al2012}, 21--%
             \citet{ablimit_et_al2020}, 22--\citet{bird_et_al2022}
             (from data of K-type giants), 23--%
             \citet{yuan_zhou_et_al2023}, 24--%
             \citet{bhattacharjee_et_al2014}, 25--%
             \citet{jianling_wang_et_al2022}, 26--\citet{bird_et_al2022}
             (from data of BHB stars), 27--%
             \citet{guang_chen_sun_et_al2023}, 28--%
             \citet{bajkova_bobylev2016}, 29--\citet{y_huang_et_al2016},
             30--\citet{eadie_et_al2015}, 31--%
             \citet{patel_et_al2018}, 32--%
             \citet{ibata_et_al2024} (within 50\,kpc), and 33--%
             \citet{ibata_et_al2024} (estimate of virial mass within 
             200\,kpc).}
    \label{fig:bobylev_baykova2023figure2adapt}
  \end{figure}

  Figure \ref{fig:bobylev_baykova2023figure2adapt} shows our new 
  estimate of the enclosed mass of the Milky Way within 100\,kpc in 
  comparison to other estimates of the enclosed mass at various radii 
  between 10\,kpc and 270\,kpc. Our results are in line with the 
  general trend of the literature.

  \subsection{NGC\,2419}\label{subsec:n2419}
  
  Finally, we use our new dataset to investigate more closely the
  proposed connection between the trailing arm of Sagittarius stream 
  and the globular cluster NGC\,2419
  \citep[e.\,g.,][]{irwin1999, newberg_et_al2003, ruhland_et_al2011,
                    belokurov_et_al2014, sangmo_tony_sohn_et_al2018,
                    massari_et_al2019, bellazzini_et_al2020, 
                    davies_et_al2024, 
                    chen_gnedin2024, rostami_shirazi_et_al2024}.
  The (mean) position of NGC\,2419 in 
  \autoref{fig:dist_sgr_stream_vel_struct_main} was found to be
  consistent with the approaching group of BHB stars at the apocentre 
  of the trailing arm.      

\begin{figure}
    \centering
    \includegraphics{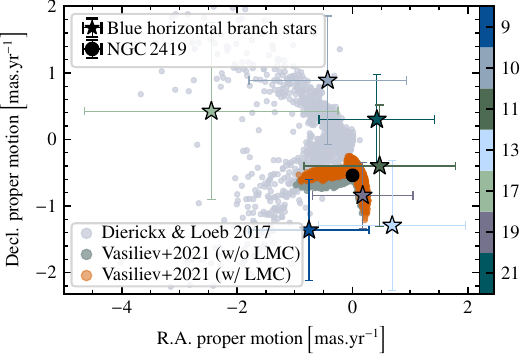}
    \caption{Proper motion of the identified BHB stars (stars) in
             comparison to the simulation data by
             \citet{dierickx_loeb2017} (point markers) and
             \citet{vasiliev_et_al2021} (square and plus markers showing
             rescaled proper motions by 1/$f$ where $f$ is the same
             factor used to rescale the heliocentric distances of the 
             simulation data, that is, $1.11\!\pm\!0.02$, and we note
             that the uncertainties in the rescaled proper motions
             resulting from the uncertainty in $f$ are smaller than the
             symbol size) and the globular cluster NGC\,2419 (filled
             circle). We show the predicted motion along right ascension
             and declination in the simulations across the same angular
             range of the stream as the BHB strars cover. The proper
             motion estimate of NGC\,2419 correspond to the values
             provided by \citet{vasiliev_baumgardt2021}.}
    \label{fig:proper_motions}
  \end{figure}

  Additionally, a subsample of BHB stars with estimates of proper
  motion in \textit{Gaia} DR3 was prepared. Here, we adapt the 
  procedure suggested by \citet{fabricius_et_al2021} and deselect 
  spurious astrometric solutions in \textit{Gaia} DR3 with
  $\text{IPD\_GOF\_HARMONIC\_AMPLITUDE} \!\geq\! 0.1$. This excludes
  candBHB8 and candBHB16. \autoref{fig:proper_motions} shows that the 
  remaining targets with \textit{Gaia} DR3 proper motions across right 
  ascension and declination \citep{gdr3} tentatively agree with data for
  NGC\,2419 by \citet{vasiliev_baumgardt2021} as well as with model
  predictions by \citet{dierickx_loeb2017} and
  \citet{vasiliev_et_al2021}. CandBHB9, candBHB10, candBHB11, candBHB13,
  candBHB17, candBHB19, and candBHB21 could therefore be linked to the
  apocentre of the trailing arm of the Sagittarius stream and its
  extension, the spur feature, across phase space. 
  
  \section{Conclusion}\label{sec:sum}
  
  The objective of this study is twofold: to evaluate the effectiveness
  of selecting candidate BHB stars in the outer halo beyond 
  $\sim$80\,kpc and to constrain the kinematics of the apocentre and 
  spur feature of the sparsely studied outer regions of the trailing arm
  of the Sagittarius stream. Our ESO/VLFT/FORS2 pilot programme 
  confirmed that 20 out of the 25 candidate BHB stars chosen from the 
  samples by \citet{thomas_et_al2018} and \citet{starkenburg_et_al2019} 
  are true BHB stars based on the spectroscopically estimated 
  effective temperatures and surface gravities from fitting with 
  \textsc{\lowercase{RVSpecFit}}
  (\autoref{fig:class_candidate_bhb_stars_eso_p106programs}). As a 
  consequence, we can confidently push the photometric selection 
  techniques developed in \citet{thomas_et_al2018} and
  \citet{starkenburg_et_al2019} to stars fainter than the limits of 
  \textit{Gaia} \citep[$G \!\simeq\! 21$\,mag,][]{gdr3}. The present 
  study has been one of the first attempts to examine the performance 
  of photometric selections of candidate BHB stars 
  at these distances with a dedicated, spectroscopic follow-up program. 
  
  Despite its exploratory nature, this study offers also some insight 
  into the behaviour on the VLT/FORS2 instrument and the 600B+22 grism
  we used when observing stars at high air mass (above 1.5) where 
  second order effects in the mapping from pixel to wavelength space 
  across the detector are expected due to telescope flexure under 
  gravity. Although the current study is based on a small sample of 
  candidate BHB and velocity standard stars to assess and quantify the 
  above-mentioned variation(s), the findings from adopting the same 
  methodology to estimate heliocentric LOS velocities of the velocity 
  standard stars as employed for the target stars suggest systematic 
  offsets from reference values on the order of at least 
  10--25\,km\,s$^{-1}$. These expected and corrected systematics in the 
  LOS velocity measurements of the target stars in our observational 
  setup made no significant difference to the overall conclusions that 
  we draw from the available phase space data of the BHB stars 
  (Sects. \ref{subsec:accuracy_los_vels} and 
  \ref{fig:dist_sgr_stream_vel_struct_main})
  
  The FORS2 spectra of the confirmed BHB stars provide further 
  observational constraints on the current Milky Way and Sagittarius 
  stream models
  \citep[e.\,g.,][]{dierickx_loeb2017, fardal_et_al2019, 
                    vasiliev_et_al2021},
  using the selected data on the spur feature of the Sagittarius debris.
  The velocity structure of the apocenter of the trailing arm of the
  Sagittarius stream, at heliocentric distances of
  $D_\odot \!\simeq\! 100$\,kpc, as traced by the identified BHB stars,
  is offset from the model predictions by \citet{dierickx_loeb2017} and
  \citet[][\autoref{fig:dist_sgr_stream_vel_struct_main}]{%
    vasiliev_et_al2021}.
  The investigation of heliocentric distance and LOS velocity in the 
  N-body simulation of the disruption of the Sagittarius dwarf galaxy 
  in the presence of the LMC by \citet{vasiliev_et_al2021} in 
  comparison to the data of the BHB stars has shown 
  that the observations can be described well by the model predictions 
  by \citet{vasiliev_et_al2021}, including the spur feature, if the 
  heliocentric distances of the simulation data are rescaled by a factor
  $f \!=\! 1.11^{+0.02} _ {-0.02}$. This would correspond to a change 
  in the outer slope of
  $\beta \!=\! 2.47^{+0.26}_{-0.21}%
  \left(\text{versus}\ 2.43^{+0.16}_{-0.13}\right)$
  and in turn leads to a refined estimate of the enclosed Milky Way 
  mass within the apocentre of the trailing arm slightly lower than the 
  estimate by \citet{vasiliev_et_al2021} at 
  $(5.3\!\pm\!0.4)\!\times\!10^{11}\,\mathrm{M}_\odot$.
  
  A limitation of this study is that the use of BHB stars as tracers of
  the metal-poor Milky Way halo does not permit the investigation of
  metallicities. This is due to the difficulty of constraining the
  metallicities of hot stars in general through optical spectroscopy,
  and in particular in the cases of low spectral resolution, due to the
  blending of metallicity-sensitive lines by strong Balmer lines. Future
  studies combining different stellar tracers where also estimates of
  metallicity can be obtained, will shed more light on not only the
  kinematics but also chemistry of distant substructure in the Galactic
  halo.

  The current data highlight the importance of providing a more detailed
  view of the outer Galactic halo through various standard candle stars
  that also serve as good kinematic tracers. The findings of this 
  research provide first insights  on the kinematics of the one of the
  most distant extensions of the Sagittarius stream currently known, 
  the spur feature. We expect that the insights gained from this study 
  will provide helpful constraints to modellers of the Sagittarius 
  stream and the Galactic potential.

  \begin{acknowledgements}
    We want to express our gratitude to Eugene Vasiliev for providing us
    the uncertainties of the scale radius, inner and outer slope, and
    transition steepness of the fiducial model of the halo density
    profile in \citet{vasiliev_et_al2021}. MB would also like to thank 
    Sergey E. Koposov for their help in understanding and using some of 
    the new features in \textsc{\lowercase{RVSpecFit}}
    \citep{koposov_et_al2011, koposov2019}. MB would like to extend 
    his thanks to Samuel Rusterucci for their help in
    proofreading the paper draft. MB and ES acknowledge funding through 
    VIDI grant ``Pushing Galactic Archaeology to its limits" (with 
    project number VI.Vidi.193.093) which is funded by the Dutch 
    Research Council (NWO). This research has been partially funded 
    from a Spinoza award by NWO (SPI 78-411). GFT and EFA acknowledge 
    support from the Agencia Estatal de Investigaci\'on del Ministerio 
    de Ciencia en Innovaci\'on (AEI-MICIN) and the European Regional 
    Development Fund (ERDF) under grant number 
    PID2020-118778GB-I00/10.13039/501100011033 and the AEI under grant 
    number CEX2019-000920-S. EFA acknowledges support from HORIZON TMA 
    MSCA Postdoctoral Fellowships Project TEMPOS, number 101066193, 
    call HORIZON-MSCA-2021-PF-01, by the European Research Executive 
    Agency. AB acknowledges the Edinburgh Doctoral College Scholarship 
    that funded her work. GEM acknowledges financial support from 
    Natural Sciences and Engineering Research Council of Canada (NSERC) 
    through grant RGPIN-2022-04794, and from an Arts \&
    Sciences Postdoctoral Fellowship at the University of Toronto. This
    research has made use of the VizieR catalogue access tool, CDS,
    Strasbourg, France (DOI : 10.26093/cds/vizier). The original
    description of the VizieR service was published in 2000, A\&AS 143,
    23. This work is based on data obtained as part of the Canada-France
    Imaging Survey, a CFHT large programme of the National Research
    Council of Canada and the French Centre National de la Recherche
    Scientifique. Based on observations obtained with MegaPrime/MegaCam,
    a joint project of CFHT and CEA Saclay, at the Canada-France-Hawaii
    Telescope (CFHT) which is operated by the National Research Council
    (NRC) of Canada, the Institut National des Science de l'Univers
    (INSU) of the Centre National de la Recherche Scientifique (CNRS) of
    France, and the University of Hawaii. This research was supported by
    the International Space Science Institute (ISSI) in Bern, through
    ISSI International Team project 540 (The Early Milky Way). We
    additionally made use of \textsc{\lowercase{adjustText}}, version
    1.2.0 \citep{ilya_flyamer_2024_12570011}, Action-based Galaxy
    Modelling Architecture code
    \citep[\textsc{\lowercase{AGAMA}},][]{vasiliev2018, vasiliev2019a},
    \textsc{\lowercase{astroquery}} \citep{ginsburg_et_al2019},
    \textsc{\lowercase{Astropy}} \citep{astropy2013, astropy2018,
                                        astropy2022},
    \textsc{\lowercase{dustmaps}} \citep{green2018},
    \textsc{\lowercase{emcee}} \citep{foreman-mackey_et_al2013},
    \textsc{\lowercase{EsoReflex}} \citep{freudling_et_al2013},
    \textsc{\lowercase{Gala}} \citep{gala},
    Image Reduction and Analysis Facility
    \citep[\textsc{\lowercase{IRAF}},][]{tody1986, tody1993, iraf},
    \textsc{\lowercase{Matplotlib}} \citep{hunter2007},
    \textsc{\lowercase{NumPy}} \citep{numpy},
    \textsc{\lowercase{RVSpecFit}}
    \citep{koposov_et_al2011, koposov2019},
    \textsc{\lowercase{Scikit-learn}} \citep{scikit-learn},
    \textsc{\lowercase{SciPy}} \citep{scipy}, 
    \textsc{\lowercase{seaborn}} \citep{waskom2021}, and  
    \textsc{\lowercase{vaex}} \citep{breddels_veljanoski2018}. We
    would like to acknowledge the use of DeepL Write, a artificial
    intelligence writing assistant developed by DeepL, which was
    employed to enhance the readability and style of this article.
  \end{acknowledgements}

  \bibliographystyle{aa}
  \bibliography{refs}

  \begin{appendix}
    \section{Reduction and preparation of observed and synthetic data}%
    \label{appendix_sec:red_calib}
    
    The systematic process of extracting the spectra from the 
    two-dimensional data recorded by the MIT detector on the VLT/FORS2 
    instrument consists of four elements or components: (1) a basic 
    correction for the artifacts and `fingerprints' of the 
    two-dimensional chip data, including bias level, flat field 
    response, bad pixels, and slit illumination based on calibration 
    data gathered at daytime (bias frames, flat fields) (2) a 
    two-dimensional mapping (along the slit axis) of the dispersion 
    axis to wavelengths with measured lines of known wavelength in 
    arc-lamps at daytime and at zenith, (3) a flux calibration and 
    instrument response correction from a response curve estimated from 
    a measured spectrum in a similar setup as the science observations 
    of a flux standard star (usually within $\pm\! 3$ days of the 
    observations of the science targets) with known continuum behaviour 
    and fluxes, and (4) tracing the science spectrum along the slit in 
    the chip data while adding together the measured pixel values at 
    each mapped column within a defined aperture and subtracting the 
    median sky value outside of the aperture along each CCD column.
    The calibration of the two-dimensional wavelength solution for the
    FORS2 observations with the 600B+22 grism included fitting a
    polynomial of fourth order with residuals typically of the order of
    0.25\,\AA\ with a few outlier pixels at up to 1.00\,\AA, ensuring a
    wavelength calibration accuracy within typically 0.25\,\AA\  across
    the entire detector, allowing for an accurate representation of the
    spectral dispersion across the detector.
    
\begin{figure*}
      \includegraphics{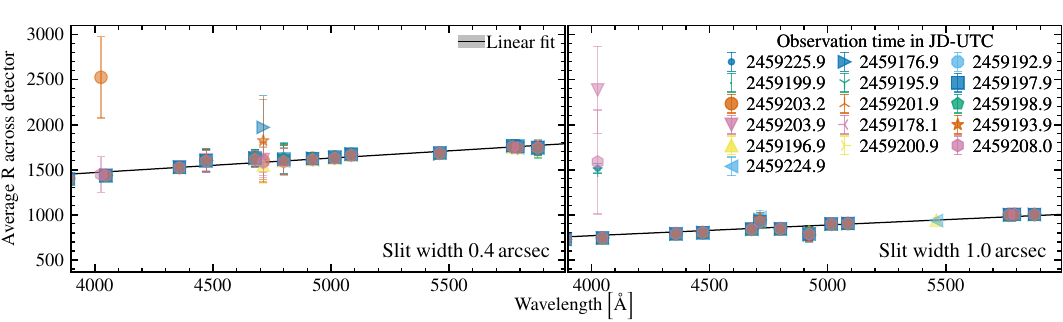}
      \caption{VLT/FORS2/600B+22 grism resolving power trends across
               detector for gathered data of arc-lamps at 
               different observation times for the two used slit 
               configurations.}
      \label{fig:vlt_fors2_600b22_resolving_power_trend}
    \end{figure*}

    With respect to the behaviour of the spectral resolution across the
    detector for the 2D data taken with the slit with width of
    1.0\,arcsec (candidate BHB stars) and 0.4\,arcsec (velocity 
    standard stars), the version of the FORS2 pipeline within the 
    \textsc{\lowercase{EsoReflex}} that we employed, also provides 
    an estimate of the average resolving power across the detector as
    measured from lines in the spectra of the arc-lamps across the 
    dispersion axis for each wavelength pixel. From
    \autoref{fig:vlt_fors2_600b22_resolving_power_trend} we can see that
    a line provides a sufficient description of the trends of the mean
    resolving power across the detector for each wavelength bin for 
    both datasets and also various observation times together as we do
    not see any significant discrepancy in the trend between 
    observation times (different markers in 
    \autoref{fig:vlt_fors2_600b22_resolving_power_trend}). Utilising a 
    $\chi^2$-minimisation using \textsc{\lowercase{NumPy.polyfit}} 
    \citep{numpy}, we found the lines that best fit the observed trends 
    in \autoref{fig:vlt_fors2_600b22_resolving_power_trend} to be
    \begin{equation}
        R \!=\! 826\!+\!0.161\,\AA^{-1}\!\times\!\lambda
        \label{equ:res_pow_lam_trend_fors2_specs_vel_sts}
    \end{equation}
    for $\text{slit\ width} \!=\! 0.4$\,arcsec (left half of 
    \autoref{fig:vlt_fors2_600b22_resolving_power_trend}) and
    \begin{equation}
        R \!=\! 293\!+\!0.119\,\AA^{-1}\!\times\!\lambda
        \label{equ:res_pow_lam_trend_fors2_specs_cand_bhb_stars}
    \end{equation}
    as a function of wavelength $\lambda$ in \AA\ for
    $\text{slit\ width} \!=\! 1.0$\,arcsec (right half of
    \autoref{fig:vlt_fors2_600b22_resolving_power_trend}).
    
    For the first observation of candBHB1 listed in 
    \autoref{tab:cand_bhb_stars} some calibration data--such as a 
    normalised flat-field image in connection with the observations of 
    a standard star to calibrate the flux of the spectrum of 
    candBHB1--were missing and we were therefore not able to run the 
    full data reduction and calibration of this observation with the 
    pipeline. However, the first spectrum of candBHB1 was automatically 
    reduced and calibrated by the ESO Science Data Quality Group in a 
    very similar setup as we adopted using the FORS2 pipeline 5.6.4 and 
    we have adopted their results for this observation.

    Regarding the synthetic spectra that we use to model the FORS2
    spectra with \textsc{\lowercase{RVSpecFit}}, for the first
    instrumental setup connected to the spectra of the candidate BHB 
    stars the synthetic stellar spectra of the PHOENIX library are re-%
    binned to the dispersion axis of the FORS2 spectra between
    3,900.94\,\AA\ and 5,989.18\,\AA\ (the wavelengths of the first and 
    last wavelength bin after 3,900\,\AA\ and before 5,990\,\AA\, 
    respectively, that we consider to be the reliable range of the
    spectra) with a $\text{linear\ dispersion} \!=\! 1.32$\,\AA\, and
    the spectral resolution was degraded from the initial
    $R \!=\! 100000$ to
    \autoref{equ:res_pow_lam_trend_fors2_specs_cand_bhb_stars}.
    All wavelengths are measured in air in this setup.
  
    Likewise, the library of synthetic stellar spectra that are used
    to fit the spectra of the velocity standard stars are prepared in
    the same way as the spectra of the candidate BHB stars with lower 
    resolution except we have to adopt the synthetic spectra in a
    higher spectral resolution that behaves as
    \autoref{equ:res_pow_lam_trend_fors2_specs_vel_sts}.
  
    Finally, the third grid of prepared synthetic stellar spectra that
    we use for the SDSS spectra have a linear dispersion of 1\,\AA\ 
    across 3,500--9,500\,\AA\ and have a uniform resolving power of
    2,000 across the dispersion axis (wavelengths are in vacuum).
  
    For all three sets of synthetic spectra we peform an initial 
    interpolation to produce a finer grid by using 0.2\,dex as step
    size for both [Fe/H] and [$\alpha$/Fe] while preserving the
    0.5\,dex sampling in $\log_{10}(g)$ and non-unform sampling in 
    effective temperature. Similar interpolations have been used in the 
    past in studies that also employ
    \textsc{\lowercase{RVSpecFit}}
    \citep[e.\,g.,][]{ting_li_et_al2019, alex_ji_et_al2021, 
                      andrew_p_cooper_et_al2023}.  
    After the first interpolation, the second interpolation is only
    performed during the fitting stage for each likelihood evaluation
    \citep[cf. with comments in][]{ting_li_et_al2019}.
    
    \section{Pan-STARRS1 $(g\!-\!r) _ 0$--effective temperature 
             relation of (candidate) Blue Horizontal Branch stars}
    \label{appendix_sec:linear_ps1gr0_color_idx_eff_temp_rel}
             
\begin{figure}
      \includegraphics{%
          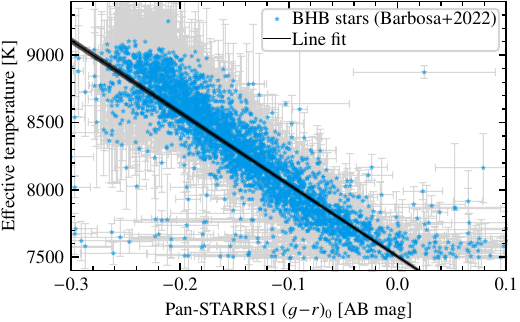}
      \caption{Correlation of $(g\!-\!r) _ {\text{PS},0}$ and effective 
               temperature among BHB stars in the
               \citet{barbosa_et_al2022} sample
               (probability of being BHB star of 100\,per\,cent)}
      \label{fig:linear_ps1gr0color_idx_eff_temp_rel}
    \end{figure}

    Gaussian priors of effective temperature were selected for the 
    posterior calculation given the spectra of the candidate BHB stars
    in the sample. For the estimation of effective temperature of each
    candidate BHB star in the sample, a relation between PS1
    $(g\!-\!r) _ 0$ and effective temperature for the BHB stars in the
    \citet{barbosa_et_al2022} sample (probability of being BHB star of
    100\,per\,cent) was derived. We adopt PS1 DR2 mean forced photometry
    \citep{ps1}. Estimates of effective temperature from Sloan Extension
    of Galactic Understanding and Exploration (SEGUE) Stellar Parameter
    Pipeline \citep{young_sun_lee_et_al2008a, young_sun_lee_et_al2008b} 
    based on SEGUE data \citep{segue} of BHB stars were chosen because
    data on these estimates are already included in the
    \citet{barbosa_et_al2022} catalogue. An orthogonal distance 
    regression was used to find the best-fitting linear relation between
    PS1 $(g\!-\!r) _ 0$ and effective temperature for the BHB stars in
    the set released by \citet{barbosa_et_al2022} and also allow to take
    into account uncertainties for both quantities in fitting. For the
    fitting we used the \textsc{\lowercase{SciPy.odr}} package
    \citep{scipy}. The results of the correlational analysis are
    summarised in \autoref{fig:linear_ps1gr0color_idx_eff_temp_rel}.
    In \autoref{fig:linear_ps1gr0color_idx_eff_temp_rel} there is a
    clear trend of decreasing effective temperature with increasing PS1
    $(g\!-\!r) _ 0$ that can be described by
    $T_\text{eff} \!=\! -(5349\!\pm\!54)\,\text{K}%
    \!\times\!\text{PS1}\ (g\!-\!r)_0%
    \!+\!(7504\!\pm\!8)$\,K
    and residual variance of the order of 5\,K (line in
    \autoref{fig:linear_ps1gr0color_idx_eff_temp_rel}).
    
    \section{Comparison of line-of-sight velocities of velocity 
             standard stars}%
    \label{appendix_sec:los_vels_vel_sts}
             
\begin{figure}
      \includegraphics{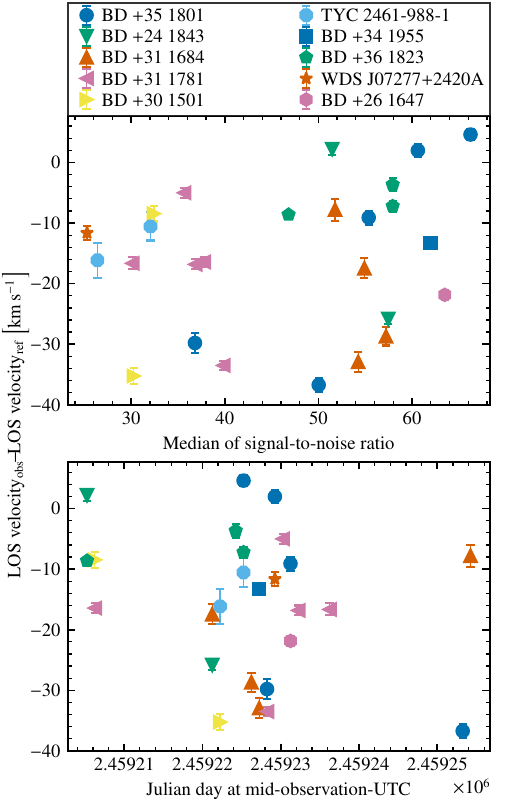}
      \caption{The results obtained from the analysis of the
               FORS2 spectra of the velocity standard stars can be
               compared to the reference LOS velocity values by 
               \citet{soubiran_et_al2018}. Times of mid-observations of 
               the spectra are shown in the abscissa of the second 
               panel.}
      \label{fig:comparison_los_vel_estimates_vel_sts_appendix}
    \end{figure}

    Our comparison of the maximum a posteriori estimates of the 
    heliocentric LOS velocities of the velocity standard stars can be
    seen in \autoref{tab:vel_sts} from fitting the optical low-%
    resolution spectra taken with the FORS2 instrument using the 
    600B+22 grism in a 0.4\,arcsec wide slit with
    \textsc{\lowercase{RVSpecFit}} and reference values by 
    \citet{soubiran_et_al2018}. Further analysis suggests a slight drop
    of the difference with higher median signal-to-noise ratio of the
    spectra between 3,900\AA\ and 5,990\AA\ (considered the reliable
    range of the spectra) in the upper panel of 
    \autoref{fig:comparison_los_vel_estimates_vel_sts_appendix} and
    no evidence of any relational association between the LOS
    velocity difference and mid-observation time (lower panel of 
    \autoref{fig:comparison_los_vel_estimates_vel_sts_appendix}). 
    Overall, these results indicate that it is sufficient to investigate
    this comparison as a function of air mass under which the stars 
    were observed as we do in \autoref{subsec:accuracy_los_vels}. 
    
    \section{Sample of candidate M-type giant stars}%
    \label{appendix_sec:cand_m_type_giants}

    A sample of (candidate) M-type giants identified by using 822,752
    low-resolution spectra from LAMOST DR9 \citep{lamost} that cover the
    spur of the Sagittarius stream \citep{jing_li_et_al2023}. 
    \citet{jing_li_et_al2023} present a catalogue of their (candidate)
    M-type giants. \citep{jing_li_et_al2023} reported 183 candidate
    members of the Sagittarius stream.   
  
    In order to compile their catalogue of (variable) RR Lyrae stars in 
    the Sagittarius stream, \citet{hernitschek_et_al2017} consider all
    RRab stars identified in \citet{sesar_et_al2017a} to be part of the 
    Sagittarius stream if they are within nine degrees of the
    Sagittarius stream orbital plane. We have adopted the same criteria
    for the selection of candidate Sagittarius M-type giant stars in 
    the \citet{jing_li_et_al2023} catalogue as we for the selection of 
    RR Lyrae in \autoref{fig:adapt_sesar_et_al2017b_figure1low_panel}. 
    The positions of all stars in the \citet{jing_li_et_al2023} 
    catalogue on the Sagittarius stream orbital plane, as defined in 
    \citet{vasiliev_et_al2021}, were calculated using the 
    \textsc{\lowercase{Gala}} \citep{gala} software. Quantitatively, 
    the \citet{hernitschek_et_al2017} selection criterion can be 
    expressed as $\left|B_\text{Sgr}\right| \!<\! 9\degr$ where
    $B _ \text{Sgr}$ is the latitude of the \citet{vasiliev_et_al2021}
    definition of the Sagittarius stream coordinate system.
  
    Several M-giants are observed in the region of the sky of the spur 
    region and consistent with its distance range. However, the 
    catalogue of \citet{jing_li_et_al2023} provides no uncertainties of 
    their photometric distances. Their distances are estimated from a 
    empirical Two Micron All Sky Survey 
    \citep[2MASS,][]{skrutskie_et_al2006} colour
    $(J\!-\!K) _ 0$-absolute magnitude $M _ {J_0}$ relation
    \citep[equation 3 in][]{jing_li_et_al2023} fitted with a subset of
    the total sample of (candidate) M-type giants closer than 4\,kpc 
    and \textit{Gaia} early DR3 inverted relative parallax 
    uncertainties \citep{gedr3} larger than 5 in the 
    \citet{bailer-jones_et_al2021} catalogue. However, in their figure 
    5, \citet{jing_li_et_al2023} examine the extent to which the 
    calculated $M _ {2\text{MASS}\ J_0}$ can be trusted and find a 
    spread of the relation of 0.64\,dex. This residual scatter of 0.64 
    will be presumably the largest component in the uncertainty of the 
    distance modulus in 2MASS \textit{J} that they use to compute the 
    photometric heliocentric distance of the (candidate) M-type giants 
    in the catalogue. 
  
    An uncertainty of 0.64\,dex would translate to a photometric 
    $\Delta D_\odot$ of at least 30\,kpc for stars at a distance of 
    $\gtrsim\! 70$\,kpc as we have in our sample. Therefore, we choose 
    not to use these stars for our purposes.
  \end{appendix}
\end{document}